\documentclass[12pt]{article}

\usepackage{tabularx} %Special tables
\usepackage{rotating} % rotate Boxes
\usepackage{epsfig}

 \usepackage{multirow} %Create multirow entries on a table
 \usepackage{graphicx} %Graphics
 \usepackage{pstricks}
 \usepackage{amssymb}

\newcommand{\gsim}{\mbox{${~\raise.25em\hbox{$>$}\kern-.70em
\lower.25em\hbox{$\sim$}~}$}}
\newcommand{\lsim}{\mbox{${~\raise.25em\hbox{$<$}\kern-.70em
\lower.25em\hbox{$\sim$}~}$}}

\author{
Enrico Nardi$^{1,2}$\footnote{enrico.nardi@lnf.infn.it},
Yosef Nir$^3$\footnote{yosef.nir@weizmann.ac.il }, 
Esteban Roulet$^4$\footnote{roulet@cab.cnea.gov.ar}$\>$    
and Juan Racker$^4$\footnote{racker@cab.cnea.gov.ar} \\[5pt] 
$^1${\normalsize
\it INFN, Laboratori Nazionali di Frascati, C.P. 13, I00044 Frascati,
Italy} \\[-2pt] 
$^2${\normalsize \it Instituto de F\'\i sica,
Universidad de Antioquia, A.A.{\it 1226}, Medell\'\i n, Colombia} \\
[-2pt] $^3${\normalsize \it Department of Particle Physics, Weizmann
Institute of Science, Rehovot 76100, Israel}\\[-2pt] 
$^4${\normalsize
\it CONICET, Centro At\'omico Bariloche, Avenida Bustillo 9500 (8400)
Argentina}}

\title{\vspace{-3cm} The importance of flavor  
in leptogenesis}

\begin{document}
\maketitle

%%%%%%%%%
\begin{abstract}
  We study leptogenesis from the out-of-equilibrium decays of the
  lightest heavy neutrino $N_1$ in the medium (low) temperature
  regime, $T\lsim 10^{12}\,$GeV ($10^{10}\,$ GeV), where the rates of
  processes mediated by the $\tau$ (and $\mu$) Yukawa coupling are non
  negligible, implying that the effects of lepton flavors must be
  taken into account. We find important quantitative and qualitative
  differences with respect to the case where flavor effects are
  ignored: (i) The cosmic baryon asymmetry can be enhanced by up to
  one order of magnitude; (ii) The sign of the asymmetry can be
  opposite to what one would predict from the sign of the total lepton
  asymmetry $\epsilon_1$; (iii) Successful leptogenesis is possible
  even with $\epsilon_1=0$. 
\end{abstract}

%%%%%%%%%%%%%%%%%%%%%%%
\section{Introduction}
A very attractive mechanism for explaining the origin of the baryon
asymmetry of the Universe ($Y_B\equiv (n_B-{\bar n}_B)/s\simeq 8.7
\times 10^{-11}$) is baryogenesis through leptogenesis
\cite{fu86,lu92}. Leptogenesis scenarios naturally appear within the
standard model minimally extended to include the see-saw mechanism
\cite{seesaw}, because all the conditions \cite{sa67} required for
generating a cosmic lepton asymmetry are generically satisfied in the
decays of the see-saw related heavy singlet neutrinos: (i) The
Majorana nature of their masses is a source of lepton number
violation; (ii) Complex Yukawa couplings induce CP violation in the
interference between the tree level and loop decay amplitudes; (iii)
For a heavy Majorana mass scale $\sim 10^{11\pm3}\,$GeV, sizable
deviations from thermal equilibrium in the primeval expanding Universe
can occur at the time the heavy neutrino decay.  Partial conversion of
the lepton asymmetry into a baryon asymmetry then proceeds by means of
anomalous $B+L$-violating electroweak sphaleron interactions
\cite{kuz85} that are standard model processes. Qualitatively it is
then almost unavoidable that a lepton (and hence a $B-L$) asymmetry is
induced in the decays of the see-saw singlet neutrinos and, since no
standard model reaction violates $B-L$, this asymmetry survives until
the present epoch. The question of whether leptogenesis is able to
explain the puzzle of the baryon asymmetry of the Universe is then a
quantitative one.

The quantitative analysis of leptogenesis has become more and more
sophisticated in recent years, taking into account many subtle but
significant ingredients, such as various washout effects
\cite{BBP0204,BP9900,pi04,pi05,ha04}, thermal corrections to particle
masses and $CP$ violating asymmetries \cite{gi04}, and spectator
processes \cite{bu01,na05}. The latter are $L$ and $B-L$ conserving
processes, such as standard model gauge interactions, some Yukawa
interactions involving the heaviest fermions, and electroweak and
strong non-perturbative `sphaleron' interactions. These processes do
not participate directly in the generation or washout of the
asymmetries (hence the name `spectator'), but have important effects
in determining the asymmetric densities of the various particles,
mainly by imposing certain relations among the chemical potentials of
different particle species \cite{bu01}.  A detailed analysis of how
spectator processes affect the washout back-reactions and concur to
determine the final value of the baryon asymmetry has been recently
presented in~\cite{na05}.

In spite of all these refinements, one potentially very significant
aspect of leptogenesis has been only rarely addressed \cite{ba00,en03,fu05}, and
that is flavor. Neglecting flavor issues is, however, justified only if
the process of leptogenesis is completed at a rather high temperature,
$T>10^{12}$ GeV. A reliable computation of the lepton and baryon
asymmetries when leptogenesis occurs in the intermediate or low
temperature windows must include flavor effects. In this paper, while
taking into account all the effects discussed in \cite{na05}, we
introduce in the analysis additional important phenomena that have to
do with the flavor composition of the lepton states involved in
leptogenesis.\footnote{In our previous work \cite{na05} we imposed
certain flavor alignment conditions, whereby the effects discussed in
this paper become irrelevant.}  In particular, we focus on the
decoherence effects that are induced by the charged lepton Yukawa
interactions on the lepton doublets produced in the decays of the
heavy neutrinos.  As soon as these Yukawa interactions approach
equilibrium, they act essentially as measuring devices that project
all the lepton densities onto the flavor basis. Lepton number
asymmetries and  washout effects then become flavor dependent, and
this can lead to a final baryon asymmetry that is different in size
and possibly even sign from the one that would arise if flavor issues
were irrelevant.

The plan of this paper is as follows. In Section \ref{sec:basic} we
present the main physics ideas that underlie the most important
effects of flavor in leptogenesis, and derive the relevant results in
a qualitative way.  In Section \ref{sec:boltzmann} we present the
network of flavor dependent Boltzmann equations, including all the
spectator processes discussed in \cite{na05}.  In Section
\ref{sec:equilibrium} we separately analyze each of the relevant
temperature regimes: we impose the appropriate equilibrium conditions
and discuss their implications for the network of Boltzmann
equations. We also present results for a set of representative flavor
structures and for different temperature regimes. In Section
\ref{sec:discussion} we compare our results to what is obtained when
flavor effects are neglected or irrelevant, and we explain the main
mechanism underlying the large enhancements of $B-L$ that flavor
effects can induce.  In Section \ref{sec:conclusions} we summarize
our main results.

%%%%%%%%%%%%%%%%%%%%%%%%%%%%
\section{The basic ideas}
\label{sec:basic}
In this section we study leptogenesis within temperature ranges well
below $10^{13}\,$GeV, where lepton flavor issues can have an important
impact  on the way leptogenesis is realized. A proper treatment of the
decay and scattering processes occurring in a thermal bath that
consists of a statistical mixture of various flavor states should
be carried out within a density matrix formalism, as discussed for
example in \cite{ba00}. Here we follow a simpler approach that is
based on a physically intuitive formulation of the problem, and
allows us to obtain all the qualitative features of
the possible solutions. The results we derive here agree well with
what will be obtained in Section~\ref{sec:equilibrium} by solving
numerically the detailed flavor-dependent Boltzmann equations that are
derived in Section~ \ref{sec:boltzmann}.

%%%%%%%%%%%%%%%%%%%%%%%%
% \setcounter{subsection}{1}
 \subsection{Flavor $CP$ violating effects}
 \label{subs:flavorCP}
The heavy singlet neutrinos $N_\alpha$ needed for the see-saw model
decay into lepton and antilepton doublets, implying lepton number
violation. We denote the lepton doublets produced in the $N_\alpha$
decays with the same index $\alpha=1,\,2,\,3$:\footnote{
  The subindex in $\ell_{1,2,3}$ bears no relation to the neutrino mass
  eigenstates $\nu_{1,2,3}$.}
\begin{eqnarray}
  \label{eq:1}
\nonumber \Gamma_\alpha &\equiv& \Gamma(N_\alpha\to \ell_\alpha\, H),
\\ \nonumber \bar \Gamma_\alpha &\equiv& \Gamma(N_\alpha\to \bar
\ell^\prime_\alpha\, \bar H).
\end{eqnarray}
$CP$-violation in $N_\alpha$ decays can manifest itself in two
different ways:\\
1. Leptons and antileptons are produced at different rates,
\begin{equation}
  \label{eq:2a}
  \ \Gamma_\alpha \neq \bar\Gamma_\alpha.
  \end{equation}
2. The leptons and antileptons produced in $N_\alpha$ decays
are not $CP$ conjugate states,
\begin{equation}
  \label{eq:2b}
\ CP(\bar\ell^\prime_\alpha) \equiv  \ell^\prime_\alpha \neq  \ell_\alpha.
\end{equation}
If the rate of charged lepton Yukawa interactions is much slower than
the rate of the heavy neutrino ($N-\ell$) Yukawa interactions, flavor
issues can be neglected since, regardless of their flavor composition,
$\ell_\alpha$ and $\bar\ell^\prime_\alpha$ remain coherent states
between two successive interactions. In this case only the effect in
eq.~(\ref{eq:2a}) is important. Indeed, leptogenesis studies have
generally concentrated on this first effect. The situation is,
however, different if leptogenesis occurs when the processes mediated
by the $\tau$ (and possibly $\mu$) charged lepton Yukawa couplings are
faster than the $N-\ell$ Yukawa interactions (a sufficient condition
for this is that these processes occur at a rate comparable to the
expansion rate of the Universe). Then, $\ell_\alpha$ and
$\bar\ell^\prime_\alpha$ are no longer the interacting states
populating the thermal bath and $\ell_i$ and $\bar \ell_i$ with
$i=\tau,(\mu,e)$ should be considered instead. The second effect in
(\ref{eq:2b}) can then become of major importance for the generation
of cosmic asymmetries.

The amount of lepton asymmetry produced per $N_\alpha$ decay ($\alpha
=1,\,2,\,3$) is 
\begin{equation}
  \label{eq:3}
  \epsilon_\alpha
  =\frac{\Gamma_\alpha-\bar\Gamma_\alpha}{\Gamma_\alpha+\bar\Gamma_\alpha}. 
\end{equation}
In order that non-vanishing $\epsilon_\alpha$ asymmetry would arise,
the two sets of three states $\ell_\alpha$ and $\ell^\prime_\alpha$
cannot form an orthogonal basis because, to get $CP$ violation from
loops, it is required that the lepton doublet coupled to the external
$N_\alpha$ (the decaying heavy neutrino) couples also to a virtual
$N_\beta$ (with $\beta\neq\alpha$) appearing in the loop. Assuming a
hierarchical pattern for the heavy $N_\alpha$ masses $M_\alpha$, and
given the non-orthogonality of the $\ell^{(\prime)}_{1,2,3}$ states,
the most natural situation is then that any asymmetry produced in
$N_{2,3}$ decays is quickly erased by fast $L$ violating interactions
involving $N_1$. One can still envisage a situation in which $\ell_3
\perp \!\!\!\!\!/ \ell_2$ is responsible for $\epsilon_2\neq0$, but
an approximate orthogonality $\ell_2\perp \ell_1$ prevents
$\epsilon_2$ from being washed out by the $N_1$ lepton number
violating processes. In this case a lepton asymmetry produced
in $N_2$ decays can survive until all the $L$ violating processes
involving $N_1$ freeze-out.  However, if this freeze-out occurs after
lepton flavor dynamics has become important, then all the
$\ell^{(\prime)}_\alpha$ will be effectively projected onto
$\ell_{\tau,(\mu,e)}$ flavor states. Then additional `flavor alignment'
conditions are necessary in order to preserve the $\epsilon_2$
asymmetry, as for example $\ell_1 \propto \ell_e$ and $\ell_{2,3}\perp
\ell_e$.  Models that realize this scenario have been studied for
example in \cite{ba00,diba05,vi05}.  In the following we disregard this
possibility, and concentrate on the decay of the lightest heavy
neutrino $N_1$ (and on $\epsilon_1$) as the dominant source of a
cosmic lepton asymmetry.

If during leptogenesis the (required) deviations from thermal
equilibrium are not large, inverse decays and other washout process
become important to determine the amount of $\epsilon_1$ asymmetry
that can be effectively converted into a baryon asymmetry by the $B+L$
violating electroweak sphaleron processes.  In the cases where the
$N_1$ heavy neutrinos have an initial thermal abundance, or when
thermal abundance is reached due to inverse decays and other
scattering processes, it is customary to express the present density
of cosmic baryon asymmetry relative to the entropy density $s$ as
follows: 
\begin{equation}
  \label{eq:4}
\frac{n_B}{s}=- \kappa_s\,\epsilon_1\, \eta.
  \end{equation}
The numerical factor $\kappa_s\simeq1.38\times 10^{-3}$
accounts for the $B-L$ entropy dilution from the leptogenesis
temperature down to the electroweak breaking scale, as well as for
the electroweak sphalerons $B-L \to B$ conversion factors. The factor 
$\eta$, that can range between 0 and 1, is the so called efficiency
(or washout) factor and accounts for the fraction of lepton
asymmetry surviving the washout processes.  If $N_1$ decays occur
strongly out of equilibrium, then all back-reactions are negligible
and $\eta\approx 1$. On the other hand, values of $\eta \sim
10^{-2}-10^{-3}$ that are typical of strong washout regimes, when
deviations from thermal equilibrium are mild, can still yield
successful leptogenesis while ensuring at the same time that $n_B/s$
is largely independent of initial conditions.
  
When $M_1\gg 10^{12}\,$GeV, successful leptogenesis requires that the
$N_1-\ell_1$ couplings are sizable and, in particular, larger than
all the charged lepton Yukawa couplings. Then in the relevant
temperature window, around $T \approx M_1$, charged lepton Yukawa
processes are much slower than the processes involving $N_1$ (and also 
slower than the rate of the Universe expansion). In this regime, the
composition of the two states $\ell_1$ and
$\bar\ell^\prime_1$ in terms of the lepton flavor states $\ell_i$
($i=\tau,\,\mu,\,e$) is irrelevant since the doublet states
produced in the decays keep coherence between two different 
scatterings involving $N_1$. However, if leptogenesis occurs at lower
temperatures, processes mediated by the $\tau$ Yukawa coupling (and for
$T\lsim 10^{9}\,$GeV also the processes mediated by the $\mu$ Yukawa
coupling) become faster than the reactions involving $N_1$. Then $\ell_1$
and $\bar\ell^\prime_1$ lose their coherence between two subsequent
$L$-violating interactions. Before they can rescatter in reactions
involving $N_1$, they are projected onto the lepton flavor basis with
respective probabilities for each flavor $i$:
  \begin{equation}
    \label{eq:5}
    K_{1i} = |\langle \ell_1 | \ell_i   \rangle|^2
    \qquad {\rm and} \qquad 
    \bar K_{1i} = |\langle \bar \ell^\prime_1 | \bar \ell_i
\rangle|^2 \qquad\qquad (CP(\bar\ell_i)=\ell_i).
      \end{equation}
In the following, we drop the index $1$ in $K_{1i},\,\bar K_{1i}$ (as
well as in the decay rates $\Gamma_1,\,\bar\Gamma_1$), leaving understood
that we always refer to $N_1$ related quantities. We will also
concentrate exclusively on the medium and low temperature regimes, 
in which the charged lepton Yukawa interactions effectively `measure',
at least in part, the flavor composition of $\ell^{(\prime)}_1$. We
distinguish the following possibilities:
\begin{enumerate}
\item Alignment: if the $i=\tau$ (or $i=\tau,\mu$) Yukawa processes are in
equilibrium, but both $\ell_1$ and $\ell_1^\prime$ are aligned or
orthogonal to one flavor $i$ ($K_i = \bar K_i = 1$ or 0), leptogenesis
reduces to a one-flavor problem, much like the unflavored case of the
high temperature regimes. Different flavor alignments mainly affect
just the way in which the lepton asymmetry gets distributed between
lepton left- and right-handed states (the latter being sterile with
respect to all $L$-violating processes) and this yields numerical
effects not larger than a few tens of percent. A detailed analysis of the
various aligned cases has been recently presented in~\cite{na05}.
        
\item Non-alignment, with only the $\tau$ (but not the $\mu$ and $e$) Yukawa
processes in equilibrium: the $\mu$ and $e$ flavor components of
$\ell_1$ and $\bar\ell^\prime_1$ are not disentangled from each other
during leptogenesis. It is then convenient to introduce two combinations of
the $\mu$ and $e$ flavors, $\ell_b$ and $\bar\ell^\prime_b$, that are
orthogonal to, respectively, $\ell_1$ and $\bar\ell_1^\prime$, thus
implying $K_b=\bar K_b=0$. The two combinations of $\mu$ and $e$,
$\ell_a$ and  $\bar\ell^\prime_a$, that are orthogonal to,
respectively, $\ell_b$ and $\bar\ell^\prime_b$, satisfy $K_a=1-K_\tau$
and $\bar K_a=1-\bar K_\tau$ (in general
$CP(\bar\ell^\prime_{a,b})\neq \ell_{a,b}$). In this case, even in the
absence of any particular alignment condition, leptogenesis can be
treated as an effectively two-flavor problem.
        
\item Non-alignment, and both $\tau$ and $\mu$ Yukawa interactions
fast: all the lepton flavor components of $\ell^{(\prime)}_1$ are
effectively resolved, and the thermal bath is populated by the ($CP$
conjugate) flavor states $\ell_i$ and $\bar \ell_i$ ($i=e,\mu,\tau$).
The full three-flavor problem has to be considered in this case. (The
onset of thermal equilibrium for the electron Yukawa coupling occurs
at temperatures too low to be relevant for standard leptogenesis
scenarios, but would not add qualitative changes to this picture).
        
\end{enumerate}
      
In the rest of this paper we address the lepton flavor issues for
leptogenesis in the above cases 2 and 3. We show how these issues
have a significant impact on the way leptogenesis is realized in the
intermediate and low temperature regimes.
      
%%%%%%%%%%%%%%%%%%%%%%%%%%%%%%%
%%%%%%%%%%%%%%%%%%%%%%%%%%%%%%%

\subsection{Lepton Flavor Asymmetries} 
 \label{subs:flavorAsy}
The $CP$ asymmetry for $N_1$ decays into the $\ell_j$ lepton flavor is
defined as 
\begin{equation}
  \epsilon^j_1 = \frac{ \Gamma(N_1\to \ell_j H)-
    \bar\Gamma(N_1\to \bar \ell_j \bar H)} {\Gamma + \bar\Gamma} 
     \equiv \frac{ \Gamma_j -\bar \Gamma_j }
    {\Gamma + \bar \Gamma }. 
\label{epsi} 
\end{equation}
Since, by definition, $K_j=\Gamma_j/\Gamma$ and  $\bar K_j=\bar
\Gamma_j/\bar \Gamma$, we can conveniently express $\epsilon^j_1$ as
\begin{equation}
\epsilon_1^j = \frac{ \Gamma\,K_j -\bar \Gamma\, \bar K_j }
    {\Gamma + \bar \Gamma }  
= \frac{K_j+\bar K_j}{2}\,\epsilon_1 +\frac{K_j-\bar K_j}{2} \simeq
\epsilon_1\,  K^0_j+ \frac{\Delta K_j}{2}.
\label{epsiapprox}  
\end{equation}
Here $\epsilon_1$ is the total asymmetry in $N_1$ decays defined in
eq.~(\ref{eq:3}), $\Delta K_j \equiv K_j - \bar K_j$, and $K^0_j\equiv
\Gamma^0_j/\Gamma^0$ represents the ratio of the ($CP$ conserving)
tree level decay amplitudes, with $K^0_j=\bar K^0_j$.  In the last
equality in eq.~(\ref{epsiapprox}), the first term, proportional to
$\epsilon_1$, corresponds to $CP$ violating effects of the first type
[eq.~(\ref{eq:2a})], the second term, proportional to $\Delta K_j$, is
an effect of the second type [eq.~(\ref{eq:2b})] since it vanishes
when $CP(\bar\ell^\prime_1)=\ell_1$, and higher order terms of ${\cal
O}(\Delta K_j\cdot\epsilon_1)$ are neglected.

For temperature regimes where the flavor states $\ell_j$ are the ones
relevant for leptogenesis, rather than the $\ell_\alpha$ states,
eq.~(\ref{eq:4}) should be replaced with 
\begin{equation}
  \label{eq:4flavor}
\frac{n_B}{s}=-\kappa_s\, \sum_{j=1}^{n_f}\epsilon^j_1\, \eta_j,
  \end{equation}
where $n_f=2(3)$ is the relevant number of `active' flavors in the
medium (low) temperature regime. In these regimes, off-shell $\Delta
L=2$ washout processes involving in particular $N_{2,3}$ are generally
negligible, and hence all the relevant washout processes involve just
the heavy neutrino $N_1$ and are therefore associated with the flavor
projectors $K_j$. Since, as will be discussed in more detail in
Section~\ref{sec:discussion}, the efficiency factors are, to a good
approximation, inversely proportional to the washout rates, we can
write $\eta_j\simeq\min\left({\eta}/{K^0_j},1\right)$, where $\eta$
represents the washout factor one would obtain neglecting flavor
effects. The value of $\eta_j$ saturates to unity when $K^0_j \approx
\eta$ and, in the cases we have studied numerically, this typically
occurs for $K^0_j\approx{\rm few}\times10^{-2}$. In these cases, one
can think of the decay of $N_1\to\ell_j H$ as one that proceeds much
like strongly out-of-equilibrium decays. However, if $K^0_j\ll 1$, the
condition $\sum_i K^0_i =1 $ implies that for (one or two of the)
other flavors, the back reactions are rather fast and in particular
can be quite effective in populating the $N_1$ states, so as to keep
their abundance close to thermal during a relevant part of the
leptogenesis era. As in the standard thermal leptogenesis scenarios,
this ensures independence from initial conditions and, moreover, that
a sizable amount of the asymmetry, much larger than a fraction
$K_j^0$, can end up surviving in the $\ell_j$ flavor. Inserting
eq.~(\ref{epsiapprox}) into eq.~(\ref{eq:4flavor}) gives:

\begin{equation}
  \label{eq:4eta2}
\frac{n_B}{s}\approx -\kappa_s\, \left\{ \matrix{n_f\, \epsilon_1\eta
+ \eta \sum_{i}\frac{\Delta K_i}{2 K_i}\hspace{4cm} &K^0_i \gsim \eta
{\ \rm \ for\ all\ } i \cr\cr 
\eta (n_f-1)\, \epsilon_1 +K_j\epsilon_1 % +(K^0_j-\eta)\epsilon_1
+\eta\sum_{i\neq j}\frac{\Delta K_i}{2 K_i} +\frac{ \Delta K_j}{2} & \quad\ 
K^0_j \lsim \eta, \quad K^0_{i\neq j} > \eta  }
\right.
  \end{equation}
\medskip 

We note the following:
\begin{enumerate}
\item In the second line of this equation, that corresponds to the
situation in which the value of $\eta_j$ saturates to $\approx 1$, the
first three terms are suppressed at least as $\eta$, while the last
term is not.
\item When $K^0_j\ll \eta$, an alignment condition
is approached \cite{na05}.
Flavor effects become strongly suppressed here when all $\Delta K_i$ vanish,
as is always the case in the two flavor situations (see 
section~\ref{sec:equilibrium}). 
\item If $\Delta K_i=0$ for all $i$, that is in the absence of $CP$
violating effects of the type of eq.~(\ref{eq:2b}), and if
$K^0_i>\eta$ for all $i$, that is away from alignment conditions,
then the final baryon asymmetry is always enhanced by a factor $n_f$
with respect to the case in which flavor effects are irrelevant (or
are neglected). This enhancement occurs independently of the
particular values of the $K^0_i$, as is clearly seen from 
fig.~\ref{figure1}, and it  holds also with respect to the
aligned cases discussed in \cite{na05} since alignment effectively
enforces the condition $n_f=1$. It is easy to understand the reason
for the $n_f$ enhancement. When the total decay rates are projected
onto the flavor $i$, each flavor asymmetry gets reduced compared to
the total asymmetry: $\epsilon_1^i\sim\epsilon_1 K^0_i$. However, this
is compensated by a suppression of the corresponding washout factor,
$\eta_i\sim\eta/K^0_i$. Then the sum over flavors yields the $n_f$
enhancement.

\item When $\ell_1$ has approximately equal projections onto the
different $\ell_i$ doublets, implying that $K^0_i\approx1/n_f$ for all
$i$, the result of the previous point still holds, regardless of the
particular values $\Delta K_i\neq 0$. This can be easily seen from the
first line of eq.~(\ref{eq:4eta2}) by noting that $\sum_i \Delta K_i=0$.

\item In the general situation, the terms proportional to $\Delta K_i$
do not vanish. Let us stress that $|\Delta K_i/\epsilon_1|$, being the
ratio of two (higher order) $CP$ violating quantities, is not
constrained to particularly small values and can well be sizable
bigger than unity. This means that the effects of
$CP(\bar \ell^\prime)\neq \ell$ can be the dominant ones in
determining the size of the final baryon asymmetry. Moreover, since
the signs of the $\Delta K_i$ are not directly related to the sign of
$\epsilon_1$, it is clear that one cannot infer in a model independent
way the sign of the cosmic baryon asymmetry only on the basis of
$\epsilon_1$. This situation (dominance of the $\Delta K_j$ effect) is
even more likely when one $K^0_j$ is very small and  $\eta_j \approx
1$. Then, as can be seen from the second line in eq.~(\ref{eq:4eta2}),
the term $\Delta K_j/2$ being not suppressed by the small value of
$\eta$ can easily dominate over all other terms.

\end{enumerate}

%%%%%%%%%%%%%%%%
\subsection{Dependence on Lagrangian parameters} 
 \label{subs:explicit}
We now proceed to express the various relevant quantities --  
$\epsilon_1,\,\epsilon_1^j$, $\Delta K_j$ and $K^0_j$ -- in terms of
the Lagrangian parameters. In the mass eigenbasis of the heavy
neutrinos $N_\alpha$ and of the charged leptons ($e_i$ denote the
$SU(2)$ singlet charged leptons), the leptonic Yukawa and mass terms
read: 
\begin{equation}
\label{yuk} 
{\cal L}_Y= -\frac{1}{2}M_\alpha \bar{N}^c_\alpha N^c_\alpha 
-(\lambda_{\alpha i}\,\overline{N}_\alpha\,  \ell_i\,  {\widetilde H}^\dagger
+h_{i}\, \overline{e}_i\, \ell_i\, H^\dagger
+{\rm h.c.}).
\end{equation}
Explicit computation of the vertex and self-energy contributions to
$\epsilon^j_1$ with this Lagrangian yields \cite{co96}:
\begin{equation}
        \label{eq:6} 
\epsilon^j_1=\frac{-1}{8\pi (\lambda \lambda^\dagger)_{11}}
\sum_{\beta\neq 1}
   {\rm Im}\left\{\lambda_{\beta j}\lambda^*_{1j} \left[\frac{3}{2\sqrt{x_\beta}}
       (\lambda\lambda^\dagger)_{\beta1} +\frac{1}{x_\beta} (\lambda\lambda^\dagger
       )_{1\beta}\right]\right\}, 
\end{equation}
where $x_\beta=M^2_\beta/M^2_1$.  The coefficient of
$(\lambda\lambda^\dagger)_{\beta1}$ within square brackets in
eq.~(\ref{eq:6}) comes from the leading term in the expansion for $x_\beta
\gg1$ of
\begin{equation}
  \label{eq:7}  \frac{\sqrt{x_\beta}}{1-x_\beta} +
  \sqrt{x_\beta}\left(1-(1+x_\beta)\ln
    \frac{1+x_\beta}{x_\beta}\right) = 
-\frac{3}{2}x^{-1/2}_\beta-\frac{5}{6}x^{-3/2}+{\cal O}(x^{-5/2}).
\end{equation}
The first term in the l.h.s. originates from self-energy type of
loop corrections, while the second term corresponds to the proper
vertex correction.  The coefficient of the second term
$(\lambda\lambda^\dagger)_{1\beta}$ in square brackets in (\ref{eq:6})
comes from the self-energy type of diagram with `inverted' direction
of the fermion line in the loop, and corresponds to the leading term
in the expansion $(1-x_\beta)^{-1} = -x_\beta^{-1}+{\cal
O}(x^{-2}_\beta)$.  Eq.~(\ref{eq:6}) holds when the mass splittings
between the heavy neutrino masses are much larger than the decay widths,
$M_\beta-M_1 \gg \Gamma_{\beta,1}$. The lowest order expression for the
flavor projectors reads:
\begin{equation}
K^0_j=\bar K^0_j=\frac{\lambda_{1j}\lambda^*_{1j}}{(\lambda\lambda^\dagger)_{11}}
\label{eq:K0}.
\end{equation}
By summing eq.~(\ref{eq:6}) over the flavor index we obtain the total 
asymmetry:
\begin{equation}
        \label{eq:8} 
\epsilon_1=\frac{-3}{16\pi (\lambda \lambda^\dagger)_{11}}
\sum_{\beta\neq 1}
   {\rm Im}\left\{\frac{1}{\sqrt{x_\beta}}
       (\lambda\lambda^\dagger)_{\beta1}^2 \right\}. 
\end{equation}
We do not need to separately calculate  $\Delta K_j$ in terms 
of the Lagrangian parameters since, from eq.(\ref{epsiapprox}),
$\Delta K_j/2= \epsilon_1^j-\epsilon_1\, K^0_j$. As was first noted
in ref.~\cite{co96}, the term $(\lambda\lambda^\dagger)_{1\beta}$
contributes to the flavor asymmetry $\epsilon_1^j$ in
eq.~(\ref{eq:6}), but -- since $(\lambda\lambda^\dagger)_{\beta 1}
(\lambda\lambda^\dagger)_{1\beta}$ is real -- it does not contribute
to the total asymmetry $\epsilon_1$  in eq.~(\ref{eq:8}). We conclude
therefore that it corresponds to a $CP$ violating effect of the second
type.

In the basis that we are using, which is defined by diagonal matrices
for the heavy neutrino masses $M_\alpha$ and for the charged lepton
Yukawa couplings $h_j$, the $\lambda_{\alpha j}$ couplings can be
conveniently written in the Casas-Ibarra parametrization \cite{ca01}:
\begin{equation}
\label{lambda}
\lambda_{\alpha j} = \frac{1}{v}\,\left[\sqrt{M}\,R\,\sqrt{m}\,
U^\dagger\right]_{\alpha j}
\end{equation}
where $v=\langle H\rangle$ is the Higgs vacuum expectation value, 
$M={\rm diag}(M_1,M_2,M_3)$ is the diagonal matrix of the heavy masses, 
$m={\rm diag}(m_1,m_2,m_3)$ is the diagonal matrix of the light neutrino 
 masses, $R=v\, M^{-1/2}\, \lambda\,U\, m^{-1/2}$ is an orthogonal 
complex matrix $(R^T\cdot
R=I)$  and $U$ is the leptonic mixing matrix.
%$U=V_{CKM}\,{\rm diag}(e^{-i\phi_1/2},e^{-i\phi_2/2},1)$
%is a unitary matrix with $V_{CKM}$ of the usual $CKM$ form, and 
%$\phi_1,\,\phi_2$ two Majorana $CP$ violating phases.
With this parametrization, the term $(\lambda\lambda^\dagger)^2_{\beta1}$  
that controls the total asymmetry in eq.~(\ref{eq:8}) is given by 
\begin{equation}
\label{eq18}
(\lambda\lambda^\dagger)_{\beta1}^2 =  \frac{M_1 M_\beta}{v^4}
\left(\sum_i m_i R^*_{1i}R_{\beta i}\right)^2,
\end{equation}
%Note that the sum over the index $j$ that yields the total asymmetry 
%involves $U^*_{jl}U_{jk}=\delta_{lk}$ and therefore $\epsilon_1$ 
%depends only on $R$.
while the term $(\lambda_{\beta j}\lambda^*_{1j})\,
(\lambda\lambda^\dagger)_{\beta1}$, that gives the leading
contribution to $\epsilon_1^j$ in eq.~(\ref{eq:6}), is
given by 
\begin{equation}
\label{eq17}
(\lambda_{\beta j}\lambda^*_{1j})\, (\lambda\lambda^\dagger)_{\beta1}
= \frac{M_1 M_\beta}{v^4}
\left(\sum_i m_i R^*_{1i}R_{\beta i}\right)
\left(\sum_{k,l} \sqrt{m_k m_l}  R_{\beta l} R^*_{1k}
  U^*_{jl}U_{jk}\right).
\end{equation}

Eqs.~(\ref{eq18}) and (\ref{eq17}) show that, if in the basis where
the light and heavy neutrino Majorana masses are real, $R$ were a {\it
real}\ orthogonal matrix, two very intriguing consequences would
follow:
\begin{enumerate}
\item The total asymmetry would vanish, $\epsilon_1=0$, while, in
general, $\epsilon^j_1\neq 0$. This is a surprising possibility for
leptogenesis. 
\item The asymmetries $\epsilon^j_1$, which involve the imaginary part
of eq.~(\ref{eq17}), would depend only on the $CP$ violating phases
present in $U$, which (in principle) are measurable in low energy
experiments. 
\end{enumerate}
This situation would be quite different from the one usually
considered in leptogenesis scenarios, in which the final baryon
asymmetry, being proportional to $\epsilon_1$, depends through the
matrix $R$ on all the low energy and high energy parameters. It
remains to be seen if a real matrix $R$ can naturally arise
in some model.

%%%%%%%%%%%%%%%%%%%%%%%%%%%%%%%%%%%%%%%%%%%%%%%%%%%%%%%%%%%%%%
\section{The network of Boltzmann equations}
\label{sec:boltzmann}
We consider the scenario in which the heavy neutrino masses are
hierarchical, $M_1\ll M_{2,3}$, and consequently the lepton asymmetry
is generated mainly via the $CP$ and lepton number violating decays of
the lightest singlet neutrino $N_1$ to the lepton doublets $\ell_1$
and $\bar\ell^\prime_1$. The important processes involving these
states are the following (all the $\gamma$'s below denote the
thermally averaged rates):

\begin{itemize} \itemsep=1pt
\item $N_1$ decays and inverse decays, with rates $\gamma_D=\gamma(N_1
\leftrightarrow \ell_1\, H)$ and $\bar \gamma_D=\gamma(N_1
\leftrightarrow \bar\ell^\prime_1\,\bar H)$.

\item $\Delta L=1$ Higgs-mediated scattering processes with rates such
as $\gamma_{S_s} =\gamma(\ell_1\, N_1 \leftrightarrow Q_3\, \bar t)$
and $\gamma_{S_t} =\gamma(\ell_1\, Q_3 \leftrightarrow N_1\, t)$,
where $Q_3$ and $t$ are respectively the third generation quark
doublet and the top $SU(2)$ singlet, as well as those involving gauge
bosons, such as in $\ell_1 N_1\to HA$ (with $A=W^{3,\pm} $ or
$B$). $CP$ violating effects might be numerically important in these
$2\leftrightarrow 2$ scatterings \cite{pi04,pi05}. However, they do
not add qualitatively new features to the analysis.  Therefore we
neglect $CP$ violation in all these processes and accordingly we use
the tree level expression for the flavor projectors $\bar K_j,\, K_j
\simeq K^0_j$.

\item 
The $s$-channel scattering processes $\gamma_{Ns} = \gamma(\ell_1\, H
\leftrightarrow \overline{\ell^\prime}_1\, \overline{H})$ with
on-shell $N_1$ are already accounted for by decays and inverse decays.
When the lepton doublets are projected onto the flavor basis, 
subtraction of these rates to avoid double counting must be carried out 
with care. 

\item 
 The  off-shell scatterings
$\gamma^{sub}_{Ns}$, involving the (pole subtracted) $s$-channel and the
$u$-channel, as well as the $t$-channel scatterings $\gamma_{Nt} =
\gamma(\ell\, \ell \leftrightarrow \overline{H}\, \overline{H})$,
depend on all the $K^\alpha_j$ projectors in a rather complicated
way. However, since they are subdominant in the temperature ranges we
are interested in, they can be safely neglected. 

\end{itemize}

\bigskip 

In the temperature regimes in which the charged lepton Yukawa
couplings become non-negligible ($T \ll 10^{13}\,$GeV), the
corresponding interactions define a flavor basis for the lepton
doublets. Then the decay rates and scattering processes involving the
specific flavors $\ell_i$ and anti-flavors $\bar \ell_i$ have to be
considered, and the Boltzmann equations should track the evolution of
all the relevant single-flavor asymmetries.  To obtain the appropriate
set of equations one can rely on the density operator approach
discussed in \cite{ba00}.  One defines a density matrix $\rho$ as the
difference between the density matrices for the leptons and for the
antileptons, so that $\rho_{ii} \propto Y_{L_i}$, and normalizes
$\rho$ so that $\sum_i \rho_{ii} = Y_L$.  A very useful property of
$\rho$ is that the off-diagonal terms, $\rho_{ij}$ with $i\neq j$ (the
coherences) vanish whenever the Yukawa interactions of one of the
flavor states $\ell_i$ or $\ell_j$ are in thermodynamic equilibrium,
since this effectively projects the leptonic states present in the
thermal bath onto those aligned or orthogonal to the flavors
characterized by non negligible Yukawa interactions.\footnote{The
transition region between the regimes where a specific lepton Yukawa
coupling is completely negligible, and the one in which it mediates
reactions in full thermal equilibrium, that is when Yukawa reactions
for one specific flavor are approaching equilibrium, should be treated
with care, since off diagonal entries $\rho_{ij}$ in the density
matrix might not be dumped fast enough to be neglected in the flavor
dynamics \cite{ab06}.  However, within the temperature ranges in which
the lepton Yukawa reactions for each flavor are fully in equilibrium
or strongly out of equilibrium, our Boltzmann equations can be safely
applied.}  Another property that we use is that if the population of
one state vanishes, $\rho_{ii}=0$, then all the coherences $\rho_{ij}$
associated with this state also vanish (this follows from {\it e.g.}
the inequality $\rho_{ii} \rho_{jj} \geq |\rho_{ij}|^2$ ). These
properties allow one to restrict the general equation for $\rho$ to a
subset of equations for the relevant flavor diagonal directions
$\rho_{ii}=Y_{L_i}$.

Following the approach outlined in \cite{na05}, in writing down the
Boltzmann equations we account for all the particle densities that are
relevant to the washout processes. Moreover, in the evolution
equations for the lepton flavor asymmetries we also include the term
${\rm d}Y_{L_i}^{EW}/{\rm d}z$ that formally accounts for the fact
that electroweak sphalerons constitute an additional source of lepton
flavor violation. Then, for consistency, we also need to add the 
equation ${\rm d}Y_B/{\rm d}z={\rm d}Y_B^{EW}/{\rm d}z$ to account for
baryon number violation by the sphaleron processes.  Given that
sphaleron interactions preserve the three charges $\Delta_i\equiv
B/3-L_i$ associated to anomaly-free currents, it follows that
$Y_B^{EW}/3=Y^{EW}_{L_i}$.  By subtracting the equations for the
lepton flavor densities from the equation for baryon number weighted
by a suitable factor 1/3, we obtain the following network of flavor
dependent Boltzmann equations:
\begin{eqnarray}
\frac{{\rm d}Y_{N_1}}{{\rm d}z}
&=&\frac{-1}{sHz}\left(\frac{Y_{N_1}}{Y_{N_1}^{eq}}-1\right)
\left(\gamma_D+2\gamma_{Ss}+4\gamma_{St}\right),
\label{eqyn} \\ [5pt]
\frac{{\rm d}Y_{\Delta_i}}{{\rm d}z}
&=&\frac{-1}{sHz}\left\{\left[
\left(\frac{Y_{N_1}}{Y_{N_1}^{eq}}-1\right) 
\epsilon_1^i - \frac{1}{2}(y_{\ell_i}+y_H)\,K^0_i\right] \gamma_D 
\right. \nonumber \\
  &&  
\left.\!\!\!\!\!\!\!\!\!\!\!\!\!\!\!\!\!\! 
-\left[ 2y_{\ell_i}+ 
    \left(y_t-y_{Q_3}\right)
    \left( \frac{Y_{N_1}}{Y_{N_1}^{eq}}+1\right) \right]K^0_i\,\gamma_{St}
-  \left[ \frac{Y_{N_1}}{Y_{N_1}^{eq}}y_{\ell_i}
   +    y_t-y_{Q_3} \right]K^0_i\,\gamma_{Ss}
%  - \left(y_{\ell_i} +\tilde y_\ell + 2\, y_H\right)K_i\,
%  \left(\gamma_{Ns}^{sub}+\gamma_{Nt}\right)
\right\},
\label{eqyli} 
\end{eqnarray} 
where we have used the standard notation $z\equiv M_1/T$.  In these
equations, $Y_{N_1}\equiv n_{N_1}/s$ denotes the density of the
lightest heavy neutrino (with two degrees of freedom) relative to the
entropy $s$, $y_X\equiv (n_X-n_{\bar X})/n_X^{eq}$
denote the asymmetries for the various relevant species $X= \ell,\
H,\ t,\ Q_3\,$ and all the asymmetries are normalized to the
Maxwell-Boltzmann equilibrium densities. Notice that
 $Y_{\Delta_i}$ is the $\Delta_i$ number density, also
normalized to the entropy, with $Y_{L_i}=(2y_{\ell_i}+y_{e_i})Y^{eq}$.
The reaction rates are
summed over initial and final state quantum numbers, including the
gauge multiplicities. In the asymmetries $y_X$, $X=\ell,\,H$ or $Q_3$ 
label any of the two doublet components, not their sum, and hence we
normalize $y_X$ to the equilibrium densities with just one degree of
freedom. This is different from other conventions ({\it e.g.} the one 
used in \cite{gi04}) and allows us to keep the proportionality
$y_X\propto \mu_X$ in terms of the chemical potentials, with the usual
convention that {\it e.g.}  $\mu_{\ell_i}$ is the chemical potential
of each one of the two $SU(2)$ components of the doublet $\ell_i$.
As already said, the subdominant $\Delta L=2$ off-shell scatterings, and
 $CP$ violation in ($\Delta L=1$) $2\leftrightarrow 2$ processes, 
 \cite{pi04,pi05} have been neglected in eq.~(\ref{eqyli}), and we
 make two further simplifications:
\begin{enumerate}
\item In eqs.~(\ref{eqyn})--(\ref{eqyli}) and in what follows we
ignore finite temperature corrections to the particle masses and
couplings \cite{gi04}. In particular we take all equilibrium number
densities $n_X^{eq}$ equal to those of massless particles.

\item We ignore scatterings involving gauge bosons, for whose rates no
consensus has been achieved so far \cite{gi04,pi04}. They do not
introduce qualitatively new effects and no further density asymmetries
are associated to them.
\end{enumerate}

\noindent 
We would like to emphasize the following points regarding 
eq.(\ref{eqyli}):

\begin{enumerate}

\item The washout terms are controlled by the $K$-projectors. The
sources of the asymmetry, on the other hand, receive additional
contributions from the $\Delta K$'s. Note that $\Delta K_i$ is not
simply proportional to the respective $K_i$ (see
eq.(\ref{epsiapprox})). This  has important consequences: when, for at
least one flavor, $\Delta K_i/\epsilon_1\gsim K^0_i$, the results are
qualitatively different from the case where, for all the flavors,
$\Delta K_i/\epsilon_1 \ll K^0_i$. In particular, the fact that the
sign of $\Delta K_i$ can be opposite to that of the $K_i^0\epsilon_1$
term opens up the possibility that leptonic asymmetry-densities of
opposite sign are generated. 

\item Subtraction of the on-shell contributions from the $s$-channel
$N_1$ heavy neutrino exchange, that corresponds to $\Delta L= 2$
$s$-channel scatterings, has to be performed with care. The cross
sections are flavor dependent: $\gamma^{\Delta L_i=2}_{N_s}(\ell_i H
\to \bar\ell_i \bar H)$ changes $L_i$ by two units, but other
scatterings, $\gamma^{\Delta L_i=1}_{N_s}(\ell_i H \to \sum_{j \neq
i}\bar\ell_j \bar H)$ change $L_i$ by only one unit. Furthermore,
differently from the unflavored case, the $\Delta L=0$ ($\Delta
L_i=1$) channels, $\gamma^{\Delta L_i=1}_{N_s}(\ell_i H \to \sum_{j
\neq i}\ell_j H)$, together with their asymmetries, must also be taken
into account.  Moreover, through inverse processes the asymmetries in
the flavors $j\neq i$ do affect the evolution of $y_{\ell_i}$ (this is
similar to the way $y_H,\, y_{Q_3}$ and $y_t$ act in the proper
$\Delta L_i= 1$ channels in the second line of
eq.~(\ref{eqyli})). Nevertheless, we see from eq.~(\ref{eqyli}) that
the result of the subtraction agrees with what one would obtain by
naively generalizing from the case of the flavor independent equation
(see e.g. ref.~\cite{gi04}) to the flavor dependent case.

\item Consider the case $K_i=\bar K_i=1$ (and hence $K_{j\neq i}=
  \bar K_{j\neq i}=0$). Then we have $Y_{\Delta_{j\neq i}}=0$. Since
  $Y_{B-L}=\sum_k Y_{\Delta_k}$, the equation for $Y_{\Delta_i}$
  coincides (as expected) with that for $Y_{B-L}$ in the unflavored
  case, or  in the cases with flavor alignment, see eq.~(13) in
  ref.~\cite{na05}. 

\item As a consequence of the sum-rules $\sum_i K_i=\sum_i\bar K_i=1$
(that imply $\sum_i\Delta K_i=0$ and $\sum_i\epsilon_1^i=\epsilon_1$),
the equation for $Y_{B-L} = \sum_k Y_{\Delta_k}$ has a simple form also
in the general case $K_i\neq\bar K_i\neq 1$: Similarly to the
unflavored or aligned cases \cite{na05}, this equation depends on
$\epsilon_1$ as the source term, and on a single asymmetry-density
$\tilde y_\ell\equiv \sum_i K^0_i\, y_{\ell_i}$.  The weighted sum of
asymmetries, $\tilde y_\ell$, represents the effective lepton doublet
asymmetry coupled to the washout of $Y_{B-L}$. Of course, all the
complications related with the flavor structure are now hidden in
$\tilde y_\ell$ whose detailed evolution is determined by the
additional equations that still depend explicitly on $K^0_i$ and
$\epsilon_1^i$. However, as we will see, there is always a point in
the $K$-space for which $\tilde y_\ell\propto Y_{B-L}$. In this particular
situation, the equation for $Y_{B-L}$ decouples from the other
equations and can provide a simple representative one-flavor
approximation to the flavor dependent case, that still captures some
of the main effects  of flavor.

\end{enumerate}

The network of equations eq.~(\ref{eqyn}) and (\ref{eqyli}) can be
solved after the densities $y_{\ell_i}$ (or $\tilde y_\ell$), $ \ y_H$
and $y_t-y_{Q_3}$ are expressed in terms of the quantities
$Y_{\Delta_j}$ with
the help of the equilibrium conditions imposed by the fast reactions,
as described in the next section.  The value of $B-L$ at the end of
the leptogenesis era obtained by solving the Boltzmann equations
remains subsequently unaffected until the present epoch. If
electroweak sphalerons go out of equilibrium before the electroweak
phase transition, the present baryon asymmetry is given,  assuming a single
Higgs doublet, by the relation~\cite{ha90}
\begin{equation}
n_B=\frac{28}{79}n_{B-L}.
\label{bvsbml}
\end{equation}
If, instead, electroweak sphalerons remain in equilibrium until
slightly after the electroweak phase transition (as would be the case
if, as presently believed, the electroweak phase transition was not
strongly first order) the final relation between $B$ and $B-L$ would
be somewhat different \cite{la00}.

%%%%%%%%%%%%%%%%%%%%%%%%%%%%%%%%%%%%%%%%%%%%%%%%%%%%%%%
\section{The equilibrium conditions}
\label{sec:equilibrium}
In this section we discuss the equilibrium conditions that hold in the
different temperature regimes which can be relevant to study  flavor
effects in leptogenesis.  Since leptogenesis takes place during the
out of equilibrium decay of the lightest heavy right-handed neutrino
$N_1$, {\it i.e.} at temperatures $T\sim M_1$, the relevant
constraints that have to be imposed among the different particle
densities depend essentially on the value of $M_1$.  To allow for a
straightforward estimate of the importance of flavor, we chose the
relevant temperature windows as in ref.~\cite{na05}, where flavor
effects were irrelevant because of the imposition of alignment
constraints, and also in presenting the results we follow closely that 
analysis.  We use the equilibrium conditions specific of each
temperature regime to express $y_{\ell_i}$, $y_H$ and $y_t-y_{Q_3}$ in terms of the
$Y_{\Delta_j}$'s.

%%%%%%%%%%%%%%%
\subsection{General considerations}
The number density asymmetries for the particles $n_X$ entering in
eq.~(\ref{eqyli}) are related to the corresponding chemical potentials
through 
\begin{equation}
n_X-n_{\bar X}=
\frac{g_XT^3}{6}\cases{\mu_X/T&fermions,\cr 2\mu_X/T&bosons,\cr} 
\label{nvsmu.eq}
\end{equation}
where $g_X$ is the number of degrees of freedom of $X$.  For
any given temperature regime the specific set of reactions that are in
chemical equilibrium enforce algebraic relations between different chemical
potentials \cite{ha90}.
In the entire range of temperatures relevant for leptogenesis, the
interactions mediated by the top-quark Yukawa coupling $h_t$, and by
the $ SU(3)\times SU(2)\times U(1)$ gauge interactions, are always in
equilibrium. Moreover, at the intermediate-low temperatures in which
flavor effects can be important, strong QCD sphalerons are 
also in equilibrium.  This situation has the following consequences:
\begin{itemize}
\item Equilibration of the chemical potentials for the
different quark colors is guaranteed because the chemical potentials
of the gluons vanish, $\mu_g=0$.
\item Equilibration of the chemical potentials for the two members of
an $SU(2)$ doublet is guaranteed by the vanishing, above the
electroweak phase transition, of $\mu_{W^+}=-\mu_{W^-}=0$. This
condition was implicitly implemented in eq.~(\ref{eqyli}) where we used
$\mu_Q\equiv \mu_{u_L}=\mu_{d_L}$,\ $\mu_\ell\equiv
\mu_{e_L}=\mu_{\nu_L}$ and $\mu_H\equiv \mu_{H^+}=\mu_{H^0}$ to write the
particle number asymmetries directly in terms of the number densities
of the $SU(2)$ doublets.
\item Hypercharge neutrality implies
\begin{equation} \label{hyper}
\sum_i\left( \mu_{Q_i}+2\mu_{u_i}-\mu_{d_i}-\mu_{\ell_i}-\mu_{e_i}\right)+ 
2\mu_H =0\,, 
\end{equation}
where $u_i$, $d_i$ and $e_i$ denote the $SU(2)$ singlet fermions of
the $i$-th generation.
\item The equilibrium condition for the Yukawa interactions
of the top-quark $\mu_t = \mu_{Q_3}+\mu_H$ yields:
\begin{equation}
\label{tQH}
y_t-y_{Q_3}=\frac{y_H}{2}\,,
\end{equation}
where the factor 1/2 arises from the relative factor of 2 between the number
asymmetry and chemical potential for the bosons, see
eq.~(\ref{nvsmu.eq}).

\item 
Because of their larger rates, QCD sphalerons equilibration occurs at
higher temperatures than for the corresponding electroweak processes,
presumably around $T_s\sim 10^{13}$~GeV
\cite{be03,mo97,mo92}) and in any case long before equilibrium is reached 
for the $\tau$ Yukawa processes. This implies    
the additional constraint
\begin{equation}
\sum_i\left(2 \mu_{Q_i}-\mu_{u_i}-\mu_{d_i} \right)=0\,.
\label{qcdsph}
\end{equation}

\end{itemize}

The condition in eq.~(\ref{tQH}) allows one to rewrite the r.h.s. of
eq.~(\ref{eqyli}) in terms of only the two asymmetries
$y_{\ell_i}$ and $y_H$.  To express these asymmetries in terms of the
$Y_{\Delta_i}$ we  define two matrices $C^\ell$ and
$C^H$ through the relations:
\begin{equation}
y_{\ell_i}= -\sum_j C^\ell_{ij}\>\frac{Y_{\Delta_j}}{Y^{eq}}, \qquad
\qquad y_H = - \sum_j C^H_j\, \frac{Y_{\Delta_j}}{Y^{eq}}\,.
\label{AH}
\end{equation}
The matrices $C^\ell$ and $C^H$ constitute a generalization to the
 case of flavor non-alignment of the coefficients $c_\ell$ and $c_H$,
 introduced in \cite{na05}.
The numerical values of their entries are
determined by the constraints among the various chemical potentials
enforced by the fast reactions that are in equilibrium in the
temperature range ($T\sim M_1$) where the asymmetries are produced.
The Boltzmann equations in eq.~(\ref{eqyli})
can now  be rewritten as follows:
\begin{eqnarray}
\nonumber
  \frac{{\rm d}Y_{\Delta_i}}{{\rm d}z}&=&   \frac{-1}{sHz}\left\{
    \left(\frac{Y_N}{Y_N^{eq}}-1\right)\epsilon^i_1\, \gamma_D 
    +  K^0_i\sum_j \,\left[\>\frac{1}{2}\left(C^\ell_{ij}+C^H_j\right)\, \gamma_D
 \right. \right. + \\[3pt]
&&  
\hspace{-3.7cm}  \left. \left.
\phantom{\left(\frac{Y_N}{Y^eq_N}\right)}
\left(\frac{Y_N}{Y_N^{eq}}-1\right)
\left(C^\ell_{ij}\, \gamma_{Ss} + 
\frac{C^H_j}{2}\, \gamma_{St}\right) +
\left(2\,C^\ell_{ij} +  C^H_j\,\right)
\left(\gamma_{St}+\frac{1}{2}\gamma_{Ss}\right)\right]\frac{Y_{\Delta_j}}{Y^{eq}}
\right\}.
\label{eqyli2}    
\end{eqnarray}
These equations are general enough to account for all the effects of
the relevant spectator processes (Yukawa interactions, electroweak and
QCD sphalerons) as well as for a general lepton flavor structure.

The fact that the consequences of flavor cannot be easily read off from
the system of coupled equations (\ref{eqyli2})
impedes a simple comparison with the results of the unflavored
cases. We later show the results obtained by numerically integrating the
set of coupled equations. But, in order to get some insight into the
results, it is  possible to introduce an approximation to the
general equations (\ref{eqyli2}) in the form of a one-flavor equation 
for $Y_{B-L}$ that accounts quite accurately for the numerical impact
of flavor effects for two classes of models:
\begin{enumerate} 
\item Models in which $N_1$ decays with approximately equal
rates to all flavors ($K^0_i\approx 1/n_f$ for all $\ell_i$).
\item Models in which all the flavor asymmetries $\epsilon_1^i$
are dominated by the term $K^0_i\,\epsilon_1$ [see
eq.~(\ref{epsiapprox})]. 
\end{enumerate} 
From the discussion in section~\ref{subs:flavorAsy} it is clear that
both kind of models have the common feature of being sensitive only to
$CP$ violating effects of the type in eq.~(\ref{eq:2a}).  The
approximation to be discussed below captures in full this type of
effects, but it is blind to the $CP$ violating effects of the second
type in eq.~(\ref{eq:2b}), and therefore does not yield reliable
results for the cases where the asymmetries are dominated by the
effects of $\Delta K\neq 0$.

We proceed as follows. We consider particular flavor structures
leading to $K^0_i$ values that satisfy the conditions $\sum_i K^0_i
C^\ell_{ij} = \tilde c_{\ell}$, independently of the value of
$j=1,\dots n_f$. We further introduce for the Higgs asymmetry an
average coefficient $\tilde c_{H} \equiv \sum_j C^H_j/n_f$ and neglect
the terms $\delta C^H_j= C^H_j-\tilde c_H$. (In all the cases that we
consider $\delta C^H_j/\tilde c_H\lsim 15\% $). Then, for this
particular configuration, we can add the equations (\ref{eqyli2}) 
over the flavors $i$ to obtain an equation for $Y_{B-L}$ that is
independent of flavor indexes:  
\begin{eqnarray}
\nonumber \frac{{\rm d}Y_{B-L}}{{\rm d}z}&=& \frac{-1}{sHz}\left\{
  \left(\frac{Y_N}{Y_N^{eq}}-1\right)\epsilon_1\, \gamma_D +
  \frac{Y_{B-L}}{Y^{eq}} \left[\>\frac{1}{2}\left(\tilde c_\ell+\tilde
  c_H\right)\, \gamma_D \right. \right. + \\[3pt] &&
\hspace{-3.7cm}  \left. \left.
\phantom{\left(\frac{Y_N}{Y^eq_N}\right)}
\left(\frac{Y_N}{Y_N^{eq}}-1\right)
\left(\tilde c_\ell\, \gamma_{Ss} + 
\frac{\tilde c_H}{2}\, \gamma_{St}\right) +
\left(2\,\tilde c_\ell +  \tilde c_H\,\right)
\left(\gamma_{St}+\frac{1}{2}\gamma_{Ss}\right)\right]
\right\}.
\label{eqyBmL}    
\end{eqnarray}
Of course, it is useful to proceed in this way only to the extent that
this special case is representative of a more general class of flavor
configurations (corresponding {\it e.g.} to different sets of $K_i$).
As we will see, this is indeed true for the two classes of models
mentioned above. Since eq.~(\ref{eqyBmL}) is similar in form to the
one studied in ref.~\cite{na05}, where different coefficients $c_\ell$
and $c_H$ corresponding to situations of flavor alignment were introduced,
the values of $\tilde c_\ell$ and $\tilde c_H$ will give, by direct
comparison, a measure of the possible impact of flavor effects in
these models.

%%%%%%%%%%%%%%%
\subsection{Specific temperature ranges and flavor structures}
\label{subs:ranges}
We now discuss the flavor effects in the various relevant temperature
ranges. In order to clearly show the impact of these effects, we
conduct our analysis in a way that allows for a meaningful comparison
with our results in \cite{na05} in which - due to appropriate alignment
conditions -- flavor issues played no role. Thus we use the same
temperature windows (below $T\sim 10^{13}$) as in \cite{na05}, and
(obviously) impose the same equilibrium conditions.

As explained above, some of our main results can be understood from
the specific, effectively one-flavor, cases that are described by
eq.~(\ref{eqyBmL}). These cases are presented in
Table~\ref{tab:results} and can be easily compared with the
corresponding results in Table 1 of \cite{na05}. In the last column in
Table~\ref{tab:results} we give the resulting $B-L$ asymmetry. In the
fourth and fifth columns, we give the values of $\tilde c_\ell$ and
$\tilde c_H$. We remind the reader that the sum $\tilde c_\ell +
\tilde c_H$ gives a crude scaling of the overall strength of the
washout processes, while the ratio $\tilde c_H/\tilde c_\ell$ gives a
rough estimate of the relative contribution of the Higgs asymmetry to
the washout.

\begin{figure}[t!]
\centerline{\protect\hbox{
\epsfig{file=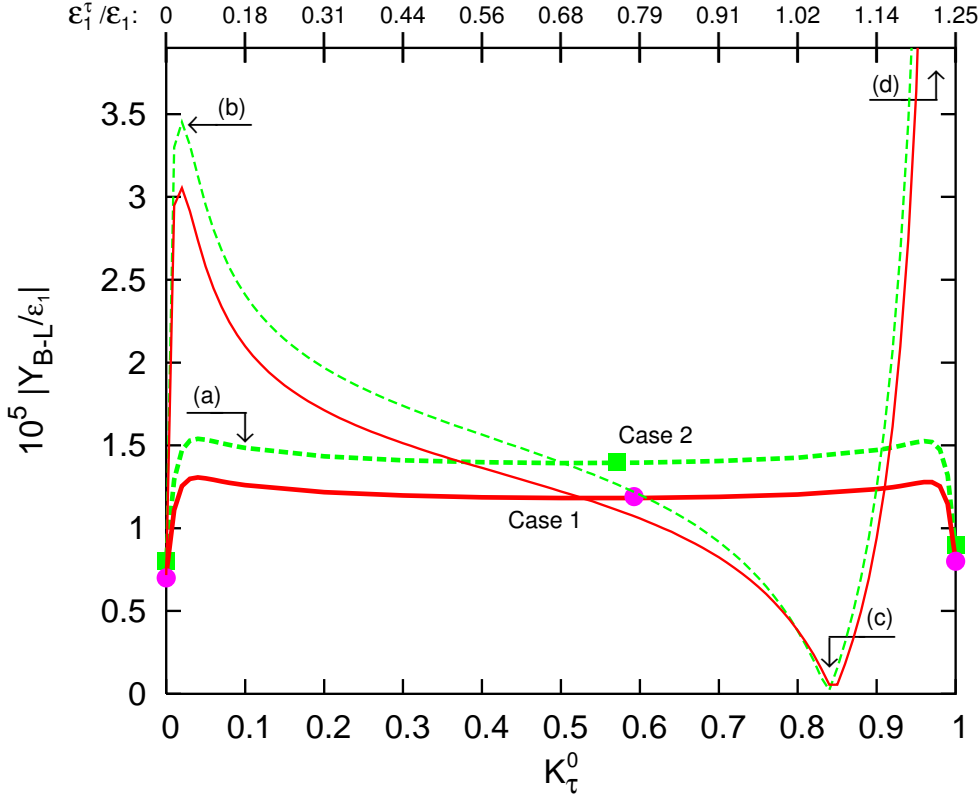 ,width=11cm,angle=270}}}
%\vskip 1.0 truecm
\caption[]{The value of the final $|Y_{B-L}|$ (in units of
  $10^{-5}|\epsilon_1|$) as a function of $K^0_\tau$ in the regimes 1
  (solid lines) and 2 (dashed lines), computed with $M_1=10^{11}$~GeV
  and $\tilde m_1=0.06$~eV.  The filled circles (regime 1) and squares
  (regime 2) give the aligned cases ($K^0_\tau=0,\,1$) discussed in  
  ref.~\cite{na05} and the non-aligned cases given in the first two rows 
  in Table~\ref{tab:results}. The thick lines correspond to
  non-aligned models for which $\Delta K_\tau=0$, implying
  $\epsilon_1^\tau/\epsilon_1=K^0_\tau$.  The thin lines give an
  example of the results for $K_\tau \neq \bar K_\tau$ assuming
  $\Delta K_\tau /2\epsilon_1 = \sqrt{K^0_\tau}/4$. The
  corresponding values of $\epsilon_1^\tau/\epsilon_1$ are marked on
  the upper $x$-axis.  The arrows with labels $(a)$,$(b)$,$(c)$ and
  $(d)$ correspond to the four panels in figure~(\ref{figure2}). Note
  that $Y_{B-L}$ changes sign in $(c)$.  }
\label{figure1}
\end{figure}

For more general flavor configurations that do not belong to the two
classes of models 1 and 2 above, and for which the approximation in
eq.~(\ref{eqyBmL}) does not hold, we present numerical results in a
graphic form in Figs.~\ref{figure1} and \ref{figure2}.  To disentangle
the impact of the various effects from that of the input parameters,
the $B-L$ asymmetry is calculated in all cases with fixed values of
$\tilde m_1=0.06$~eV and $M_1=10^{11}$~GeV, where $\tilde m_1\equiv
{v^2}(\lambda\lambda^\dagger)_{11}/{M_1}$ determines the departure
from equilibrium of the heavy neutrino $N_1$ and provides an overall
scale for the strength of the washout processes.  The value
$\tilde{m}_1=0.06$~eV is intermediate between the regime in which
departures from equilibrium are large and all washout effects are
generally negligible ($\tilde{m}_1<10^{-3}$~eV) and the regime in
which washout processes are so efficient that often the surviving
baryon asymmetry is too small ($\tilde{m}_1 \gsim 0.1$~eV). The choice
of this value is also motivated by the atmospheric neutrino
mass-squared difference if neutrino masses are hierarchical.  As
concerns $M_1$, it is clear that the relevant temperature range is
actually determined by it, yet we fix the value at $M_1=10^{11}$~GeV
in order to have a meaningful comparison of the various effects of
interest. Namely, since in each regime considered the same asymmetries
are produced in the decay of the heavy neutrinos, a comparison between
the final values of $B-L$ for the different cases can be directly
interpreted in terms of suppressions or enhancements of the washout
processes. We assume an initially vanishing value for $Y_N$ and for
all the particles density-asymmetries, but for $\tilde m_1>10^{-2}$~eV
the results are insensitive to the initial values. The values of the
parameters adopted here as well as the initial conditions are the same
as in \cite{na05}.

The four different temperature regimes that we consider are
distinguished by the additional interactions that enter into
equilibrium as the temperature of the thermal bath decreases. Of
course, the most important of these reactions will be those mediated
by the charged lepton Yukawa couplings.

%%%%%%%%%%%%%%%%%%%%%%%%%%%%%%%%%%%%%%%%%%%%%%%%%%%%%%%
%%%%%%%%%%%   TABLE %%%%%%%%%%%%%%%%%%%%%%%%%%%%%%%%%%
%%%%%%%%%%%%%%%%%%%%%%%%%%%%%%%%%%%%%%%%%%%%%%%%%%%%%%%
%
\begin{table}[t!] 
\begin{center}
  \renewcommand{\arraystretch}{1.4}
\begin{tabular}
{p{0cm}|>{\centering\small}p{1.4cm}>{\raggedleft\arraybackslash\small}p{2.3cm}>
{\raggedleft%\arraybackslash
\small}c>{\raggedleft\small}c>{\raggedright%\arraybackslash
\small}p{.4cm}c}
%HEADING 
%level 1
\omit & \multicolumn{6}{c}{\bf Equilibrium processes, constraints,
  coefficients and $B-L$ asymmetry} 
\\ \hline\hline 
%%%%%%%%%%%%%%
&&&&& \\[-14pt]
&{\small$T\,$(GeV)}& {\small Equilibrium}  & {\small Constraints} & 
$\hbox{\large $\tilde c$}_\ell$ &
$\hbox{\large $\tilde c$}_H$&
$\frac{|Y_{B-L}|}{10^{-5}\epsilon_1}$ 
 \\[4pt]  \hline \\ [-16pt]
%%%%%%%%%%%%%% III
\begin{rotate}{90}
$\!\!\!\!\!\! B=0$ 
\end{rotate}
&$10^{12\div 13}$ & \ \quad  $h_b$,\ $h_\tau \quad $ &
$
\left.
\begin{array}{l}
\raise 3pt \hbox{$b=Q_3-H,$}\\[-4pt] 
 \tau=\ell_\tau-H 
\end{array}
\right\}
 \begin{array}{l}
\qquad K^0_{\tau}\!=\!\frac{16}{27}
 \end{array}% \right.
$ &
$\begin{array}{l}\!  
\! \frac{2}{9}  
\end{array}$ &  
    $\hspace{-2mm} 
\begin{array}{l}  
\ \ \frac{7}{32}
\end{array}$ & 
$\begin{array}{l}  
1.2 
\end{array}$  
    \\[10pt] \hline 
%%%%%%%%%%%%%%%%%%%%%%%%%%%%%%%%%%%%%%%%%%%%%%%%%%%%%%%%%%% IV
 &&&&& \\[-16pt]
&$10^{11\div 12}$  &  +\quad  EW-Sph & $\sum_i(3Q_i+\ell_i)=0\ \ 
\begin{array}{l}
\quad K^0_{\tau}\!
=\!\frac{4}{7} 
\end{array} % \right.
$ &
    $\begin{array}{l}
\!\! \frac{\ 6\ }{35}  
\end{array}$ 
&  
    $\hspace{-2mm} 
    \begin{array}{l}\, 
\, \frac{97}{460} 
\end{array}$ 
& 
    $\begin{array}{l} 
  1.4  
\end{array}$  
    \\[8pt] 
%%%%%%%%%%%%%% 
\begin{rotate}{90}
$\!\!\!\!\! B\neq 0$ 
\end{rotate}
%%%%%%%%%%%%%% V
&$10^{8\div 11}$ & +\ $h_c$, $h_s$, $h_\mu$  
& 
\qquad $\left.
 \begin{array}{l}
c\!=\!Q_2\!+\!H, \\ [-3pt] 
s\!=\!Q_2\!-\!H,  \\  [-3pt] 
\mu\!=\!\ell_\mu\!-\!H
\end{array}
\right\} 
\begin{array}{l} 
\ \quad K^0_{\tau,\mu}\!=\!\frac{19}{53}  
\end{array}  % \right.
\ \ $ &                  
$\begin{array}{l}
\!\!  \frac{5}{53} % \\[-2pt]
\>
\end{array}$ &  
$\begin{array}{l}
\! \frac{47}{358}  % \\[-2pt]
\>
\end{array}$   & 
$\begin{array}{l}
\, 2.5 
\end{array}$ 
   \\[20pt] 
%%%%%%%%%%%%%% VI
& $\ll 10^{8}$ % Ref. \cite{bu01} 
& % + $h_u$, $h_d$, $h_e$  & % \multicolumn{2}{l}
 
$ \begin{array}{l}  
%\hbox{\rm EW-Sph + } \\[-4pt]
 \hbox{\rm \small  all Yukawas $h_i$ } 
 \end{array} $ 
&
$\hspace{1.5cm} % \left\{ 
\begin{array}{l}  
\qquad 
\qquad \quad \ 
K^0_{\tau,\mu,e}\!=\!\frac{1}{3} 
 \end{array}   % \right. 
$
  & 
$ \begin{array}{l}  
\frac{7}{79}   
 \end{array} $
&
$ \begin{array}{l}  
 \frac{8}{79}   
 \end{array}  $
&
$ \begin{array}{l}  
 3.0 
 \end{array}$ \\[8pt]
\hline
 \hline
\end{tabular}
\caption{\baselineskip 12pt 
The temperature window is given in the first column, the relevant
interactions in equilibrium in the second, and the constraints on the
chemical potential in the third. (Chemical potentials are labeled with
the same notation used for the fields: $\mu_{Q_i}\!=\!Q_i$,
$\mu_{\ell_i}\!=\!\ell_i$ for the $SU(2)$ doublets,
$\mu_{u_i}\!=\!u_i$, $\mu_{d_i}\!=\!d_i$, $\mu_{e_i}\!=\!e_i$ for the
singlets and $\mu_H=H$ for the Higgs.) We also give in the third
column the $K^0_i$-values that are compatible with eq.~(\ref{eqyBmL}).
The values of the coefficients $\tilde c_\ell$ and $\tilde c_H$ are
given in, respectively, the fourth and fifth column while the
resulting $B-L$ asymmetry (in units of $10^{-5}\times \epsilon_1$)
obtained for $\tilde m_1 = 0.06\,$ eV and $M_1= 10^{11}\,$GeV is given
in the last column.} 
\label{tab:results}
\end{center}
\end{table}

%%%%%%%%%%%%%%%%%%%%%%%%%%%%%%%%%%%%%%%%%%%%%%%%%%%%%%%%
%%%%%%%%%%%%%%%%%%%%%%%%%%%%%%%%%%%%%%%%%%%%%%%%%%%%%%%%
\bigskip 
\noindent
1) {\it Bottom- and tau-Yukawa interactions in equilibrium}
($10^{12}\ {\rm GeV}\lsim T\lsim 10^{13}\,$GeV). \\[3pt] 
Equilibrium for $h_b$- and $h_\tau$-interactions implies that
the asymmetries in the $SU(2)$ singlet $b$ and $e_\tau$ degrees of
freedom are populated. The corresponding chemical potentials obey the
equilibrium constraints $\mu_b=\mu_{Q_3}-\mu_H$ and
$\mu_\tau=\mu_{\ell_\tau}-\mu_H$.  Possibly $h_b$ and $h_\tau$ Yukawa
interactions enter into equilibrium at a similar temperature as the
electroweak sphalerons \cite{be03}. However, since the rate of the
non-perturbative processes is not well known, we first consider the
possibility of a regime with only gauge, QCD sphaleron and the Yukawa
interactions of the whole third family in equilibrium. 
Since electroweak sphalerons are not active, at this stage $B=0$. 

The lepton $(\ell_a,\ell_b,\ell_\tau)$ and antilepton
$(\bar\ell^\prime_a,\bar\ell^\prime_b,\bar\ell_\tau)$ flavor bases are
defined such that $\langle \ell_1 |\ell_b\rangle=\langle \bar
\ell^\prime_1 |\bar\ell^\prime_b\rangle=0$, and hence $|\langle \ell_1
| \ell_a \rangle|^2\equiv K_a = 1-K_\tau$ and $|\langle
\bar\ell^\prime_1 | \bar\ell^\prime_a \rangle|^2\equiv \bar K_a =
1-\bar K_\tau$.  The `prime' labeling the $a$ and $b$ antilepton
doublets is a reminder of the fact that in general
$\bar\ell^\prime_{a,b}$ are not the $CP$ conjugate states of
$\ell_{a,b}$. Even though the charged lepton Yukawa interactions for
both $a$ and $b$ states are negligible, the fact that $L_b
=\rho_{bb^\prime}=0$ ensures that the off-diagonal entries
$\rho_{ab^\prime}$ and $\rho_{ba^\prime}$ vanish.  Given that
$Y_{\Delta_b}/Y^{eq}=2y_{\ell_b}=0$, the set of eqs.~(\ref{eqyli2}) is
reduced to just two relevant equations, and the equilibrium conditions
(restricted to the $a,\tau$ subspace) lead to the following values for
$C^H_i$ and $C^\ell_{ij}$:
\begin{equation}
C^H = \frac{1}{16}(3 ,\>4 ) \qquad \hbox{\rm and} \qquad 
C^\ell =\frac{1}{32} \pmatrix{
 16 & 0 \cr
 1 &12 \cr}\,.
\label{case3}
\end{equation}%

The effective one-flavor approximation to non-alignment for this
regime (the first line in Table~\ref{tab:results}) corresponds to
$K^0_\tau=16/27$, as follows from the condition $\sum_i K_i\,
C^\ell_{ij}=\tilde c_\ell$ ($i,j=a,\tau$). This value ensures $\tilde
y \propto Y_{B-L}$, giving in this case $\tilde c_\ell=2/9$ (while
$\tilde c_H=\sum_{j=a,\tau}C_j^H/2=7/32$).  The enhancement of
$|Y_{B-L}|$ in our representative case is of about 60\% with respect
to the aligned cases \cite{na05} in the same temperature regime.

In Fig.~\ref{figure1} we display by a thick solid line the $B-L$
asymmetry as a function of $K^0_\tau$. This curve corresponds to  
$K_\tau =\bar K_\tau$ and is obtained by numerical integration of the
system of equations~(\ref{eqyli2}). Three special cases are denoted
with filled circles: The flavor-dependent case with $K^0_\tau=16/27$
(described in the Table) and the flavor-aligned cases $K^0_\tau=0,\,
1$ (discussed in \cite{na05}). Two features that are expected from the
qualitative discussion in section~\ref{subs:flavorAsy} are apparent in
this figure: First, the effects of flavor misalignment are quite
insensitive to the particular value of $K^0_\tau$ (as long as it
is not too close to the alignment conditions). Second, the
one-flavor approximation (the filled circle at $K^0_\tau=16/27$)
provides a good estimate of $Y_{B-L}$ for generic values of
$K^0_\tau$.

%%%%%%%%%%%%%%%%%%%%%%%%%%%%%%%%%%%%%%%%%%%%%%%%%%%
%%%%%%%%%%%%%%%%%%%%%%%%%%%%%%%%%%%%%%%%%%%%%%%%%%%
%%%%%%%%%%%%%%%%%%%%%%%%%%%%%%%%%%%%%%%%%%%%%%%%%%%

In the more general cases where $K_\tau \neq \bar K_\tau$, several
different parameters concur to determine the final value of $B-L$.
One way to explore the possible results in this case would be to
randomly sample the values of the Yukawa couplings and $CP$ violating
phases entering the expressions for $K^0_{\tau,a}$, $\Delta
K_{\tau,a}$ and $\epsilon_1$ and present the results in the form of
scatter plots. However, to gain a qualitative understanding of the
various possibilities, we proceed in a simpler way. It is based on the
fact that, if we take the tree-level $N_1\to\ell_\tau H$ decay
amplitude to zero ($\lambda_{1\tau}\to0$) while keeping the total
decay rates fixed [$(\lambda\lambda^\dagger)_{11}=\>$const.], the
$\tau$-flavor projector $K^0_\tau$ vanishes as the square of this
amplitude ($\propto|\lambda_{1\tau}|^2$) while $\Delta K_\tau$
vanishes as the amplitude ($\propto\lambda_{1\tau}$). This suggests to
adopt, as a convenient ansatz, the relation $\Delta K_\tau \propto
\sqrt{K^0_\tau}$ also for finite values of $K^0_\tau$, allowing us to
explore the interplay between the contributions of $\Delta K_\tau$ and
of $\epsilon_1 K^0_\tau$ to $\epsilon_1^\tau$ -- see
eq.~(\ref{epsiapprox}) -- as well as to $\epsilon_1^a =\epsilon_1-
\epsilon_1^\tau$, by means of a simple two-dimensional plot. To this
aim, we take $\Delta K_\tau/2\epsilon_1= \kappa_\tau \sqrt{K^0_\tau}$
and we fix $\kappa_\tau$ to the representative value of $1/4$. Thus,
for the practical purpose of carrying out a qualitative survey of the
possible different situations, we adopt the following ansatz:
\begin{equation}
\underline{\rm ansatz:} \qquad\qquad
\frac{\epsilon_1^\tau}{\epsilon_1} =
K^0_\tau+\frac{1}{4}\sqrt{K^0_\tau}.
\label{ansatz}
\end{equation}
With respect to the hierarchy between the two different $CP$ violating 
effects of eqs.~(\ref{eq:2a}) and (\ref{eq:2b}), this simple relation has
the nice property of covering all the interesting possibilities:
\begin{enumerate}
\item For $K^0_\tau \lsim 1/16$, the two terms $K^0_\tau\, \epsilon_1$ and
  $\Delta K_\tau$ are comparable in size.
\item For $1/16 \lsim K^0_\tau \lsim 3/4$, $K^0_\tau\epsilon_1 $ and
  $K^0_a\epsilon_1$ dominate the respective asymmetries.
\item For $K^0_\tau > 3/4$, we enter a regime in which $\Delta
  K_\tau/2 \gg K^0_a \epsilon_1$ and, since $\Delta K_a = -\Delta
  K_\tau$, the asymmetry $\epsilon_1^a $, that is largely dominated by
  the $\Delta K_a$ term, has now the opposite sign with respect to
  $\epsilon_1$ and $\epsilon_1^\tau$.
\end{enumerate}

In Fig.~\ref{figure1} we show with the thin continuous line the
results for this more general case. On the upper $x$-axis we have
marked for reference the value of $\epsilon_1^\tau/\epsilon_1$
corresponding to each different value of $K^0_\tau$.  The most
peculiar features are the two narrow regions marked with $(b)$ and
$(d)$ where $|Y_{B-L}|$ is strongly enhanced. The enhancement takes
place at values of $K^0_\tau$ not far from alignment. In particular,
the steep rise of $|Y_{B-L}/\epsilon_1|$ close to $K^0_\tau\approx 1$ can
reach values up to one order of magnitude larger than the vertical
scale of the figure. (Note, however, that the ansatz eq.~(\ref{ansatz})
does not yield the required behavior $\Delta K_a= -\Delta K_\tau=0$
for $K^0_a=1-K^0_\tau=0$, and therefore the thin lines in the plot
should not be extrapolated to $K_\tau\simeq 1$.) It is worth
noticing that that the two peaks correspond to values of
$Y_{B-L}/\epsilon_1$ of opposite sign, since the asymmetry changes
sign in $(c)$.  Qualitatively similar effects occur in the lower
temperature regimes discussed below, and therefore we postpone the
analysis of these results to section~\ref{sec:discussion}. As concerns
the quality of our model independent approximation, we see from
Fig.~\ref{figure1} that, in this more general case, it can provide a
reasonable estimate of $Y_{B-L}$ only for $K^0_\tau \sim K^0_a\sim
1/2$, that is when $N_1$ decays with similar rates into $\ell_\tau$
and $\ell_a$.

Given that the final value of $Y_{B-L}$ depends linearly on
$\epsilon_1^\tau$ (see eq.~\ref{eq:4flavor}), we see from
eq.~(\ref{epsiapprox}) that, for a fixed value of $K^0_\tau$,
$Y_{B-L}$ is linear in $\Delta K_\tau$.  Therefore, from the two thick
and thin lines in Fig.~\ref{figure1} that respectively give
$Y_{B-L}/\epsilon_1$ for $\Delta K_\tau=0$ and for $\Delta K_\tau/(2
\epsilon_1)=\kappa_\tau \sqrt{K^0_\tau},\,$ one can easily infer, for
each $K^0_\tau$, the value of $Y_{B-L}/\epsilon_1$ corresponding to
any other value of $\Delta K_\tau$. In particular, by reflecting the
thin line with respect to the thick one, one can figure out the
results one would obtain for $\kappa_\tau <0$.

\bigskip
\noindent
2) {\it Electroweak sphalerons in equilibrium }
($10^{11}\ {\rm GeV}\lsim T\lsim 10^{12}\,$GeV). \\[3pt]
The electroweak sphaleron processes take place at a rate per unit volume
$\Gamma/V\propto T^4\alpha_W^5\log(1/\alpha_W)$
\cite{ar98,bodeker98,ar97}, and are expected to be in equilibrium from
temperatures of about $\sim 10^{12}$~GeV, down to the electroweak
scale or below \cite{be03}. Electroweak sphalerons equilibration implies  
\begin{equation}
\label{ewsph}
\sum_i\left( 3\mu_{Q_i}+\mu_{\ell_i} \right)=0\,.
\end{equation}                                
As concerns lepton number, each electroweak sphaleron transition
creates all the doublets of the three generations, implying that
individual lepton flavor numbers are no longer conserved, regardless 
of the particular direction in flavor space along which the doublet
$\ell_1$ and $\bar \ell^\prime_1$ lie.  While the electroweak
sphalerons induce  $L_b\neq 0$, the condition $\Delta_b=0$ is not
violated, and hence eqs.~(\ref{eqyli2}) still consist of just two
equations for $Y_{\Delta_a}$ and for $Y_{\Delta_\tau}$. As concerns
baryon number, electroweak sphalerons are the only source of $B$
violation, implying that baryon number is equally distributed among
the three quark generations, that is $B/3$ in each generation. This 
modifies the detailed equilibrium conditions for the quark chemical
potentials. Besides these differences, the analysis follows closely
the one carried out in the previous regime. The coefficients $C^H_i$
and $C^\ell_{ij}$ are given by
\begin{equation} 
C^H = \frac{1}{230}(41,\> 56) \qquad \hbox{\rm and} \qquad 
C^\ell = \frac{1}{460}\pmatrix{
196 &  -24 \cr
-9 &  156 }\,. 
\label{case4}
\end{equation}
In Table \ref{tab:results} we give the values of $\tilde c_\ell$
and $\tilde c_H$ that correspond to the approximation in
eq.~({\ref{eqyBmL}). For the models described by this approximation,
flavor misalignment can induce an ${\cal O}(80\%)$ enhancement of the
final $B-L$ asymmetry compared to the aligned cases discussed in
\cite{na05}. The source of this enhancement is mainly the suppression
by a factor $\sim n_f=2$ of the washout processes that is apparent in
the reduced value of $\tilde c_\ell$. In Fig.~\ref{figure1} we give
the final value of $|Y_{B-L}|$ as a function of $K^0_\tau$ for $\Delta
K_\tau = 0$ (thick dashed line) and for $\Delta K_\tau \neq 0$ (thin
dashed line). The qualitative dependence of $Y_{B-L}$ on $K^0_\tau$ is
quite similar to regime 1: If the only $CP$ violating effects are of
the type in eq.~(\ref{eq:2a}) ($CP(\bar\ell^\prime_1)=\ell_1$), the
results are approximately independent of the particular value of
$K^0_\tau$, while relaxing this condition results again in a strong
dependence on $K_\tau^0$ and possibly strong enhancements. This
general case is again well approximated by the values of $\tilde
c_\ell$ and $\tilde c_H$ given in the Table when $K^0_\tau \sim K^0_a 
\sim 1/2$.

\bigskip
\noindent
3) {\it Second generation Yukawa interactions in equilibrium}
($10^{8}\ {\rm GeV}\lsim T\lsim 10^{11}\,$GeV). \\[3pt] 
In this regime, $h_c$, $h_s$ and, most importantly, $h_\mu$ Yukawa
interactions enter into equilibrium. Given that the electron remains
the only lepton with a negligible Yukawa coupling, the Yukawa
interactions completely define the flavor basis for the leptons as
well as for the antileptons (that are now the $CP$ conjugate states of
the leptons). Correspondingly, the lepton asymmetries are also
completely defined in the flavor basis.  In this regime, the
coefficients $C^\ell_{ij}$ and $C^H_i$ projecting the asymmetries
$(y_{\ell_e},\,y_{\ell_\mu},\,y_{\ell_\tau})$ and $y_H$ onto
$(Y_{\Delta_e},\> Y_{\Delta_\mu},\> Y_{\Delta_\tau},\>)$ are: 
\begin{equation}
C^H = \frac{1}{358}(37 ,\>52 ,\> 52) \qquad \hbox{\rm and} \qquad 
C^\ell =\frac{1}{2148} \pmatrix{
906 & -120 & -120 \cr
-75 & 688 & -28 \cr
-75 & -28 & 688}\,.
\label{case5}
\end{equation}
For the approximation of eq.~(\ref{eqyBmL}), the coefficient $\tilde
c_\ell$ in Table~\ref{tab:results} is reduced by at least a factor of
three with respect to the values of $c_\ell$ in the aligned cases
analyzed in \cite{na05}, and the final value of $Y_{B-L}$ gets
correspondingly enhanced. This is precisely the $n_f\sim 3$
enhancement expected from the qualitative discussion in
section~\ref{subs:flavorAsy}.  The estimate of $Y_{B-L}$ based on the
one-flavor approximation is, again, rather precise for the class of
models for which $\Delta K_i=0$, independently of the particular
values of $K^0_{e,\mu,\tau}$ (as long as they are not too close to 0
or 1, where approximate alignment could induce effects that suppress
$Y_{B-L}$). In the more general case with $\Delta K_{e,\mu,\tau}\neq
0$, the approximation is again reliable in the region of approximately
equal flavor composition for $\ell_1$ and $\bar\ell^\prime_1$, that is
around $K^0_e\sim K^0_\mu\sim K^0_\tau\sim 1/3$.  As regards regions
in $K$ space away from equal flavor composition, we expect that the
enhancements observed in the $n_f=2$ cases, that are especially large
when one $K^0_i$ is small (see fig.~\ref{figure1}), could be even
larger in this case since now two flavor projectors can be
simultaneously small.

\bigskip 
\noindent
4) {\it All SM Yukawa interactions and electroweak sphalerons in equilibrium}
($T\ll 10^{8}\,$GeV). \\[3pt]
Equilibration of the Yukawa processes for all quarks and leptons,
including the electron, occurs only at temperatures $T<10^6$~GeV,
which are too low to be relevant for leptogenesis (at least in the
standard scenarios). We nevertheless analyze this regime, mainly
for the purpose of comparison with analysis that assume that all
Yukawa interactions are in equilibrium, but do not include the flavor
effects \cite{bu01,na05}. 

The coefficients $C^H$ and $C^\ell$ are given by
\begin{equation}
C^H = \frac{8}{79}(1 ,\> 1,\> 1) \qquad \hbox{\rm and} \qquad 
C^\ell =\frac{1}{711} \pmatrix{
221 & -16 & -16\cr
-16 & 221 & -16 \cr
-16 & -16 & 221 }\,.
\label{case6}
\end{equation}
Given the symmetric situation of having all Yukawa interactions in
equilibrium, the point of equal flavor composition for the
$\ell_1^{(\prime)}$ doublets ($K_e=K_\mu=K_\tau=1/3$) defines a fully
flavor-symmetric situation for which
$Y_{\Delta_\tau}=Y_{\Delta_\mu}=Y_{\Delta_e}=Y_{B-L}/3$.  It is then
straightforward to see that this point corresponds to the condition
$\tilde y_\ell=\sum_i K^0_i y_{\ell_i} =- \tilde c_\ell Y_{B-L}/Y^{eq}$ that defines
our one-flavor approximation. The symmetric situation implies that an
exact proportionality is also obtained for $y_H =-\tilde c_H
Y_{B-L}/Y^{eq}$ (independent of the particular values of
$K^0_{e,\mu,\tau}$).  In agreement with the qualitative analysis in
section \ref{subs:flavorAsy}, flavor effects suppress $\tilde c_\ell$
by the large factor 3.5 ($\sim n_f$) compared to the corresponding
aligned case \cite{na05}.
%We also see that in this regime $\tilde c_H \gsim \tilde c_\ell$  and
%thus the Higgs asymmetry has the large effect of roughly doubling the
%washout effects induced by the lepton doublets alone.
In the cases restricted by $\Delta K_i=0$ for which the results of the
one-flavor approximation hold, the final value of $Y_{B-L}$ is
enhanced by a factor $\sim 2.5$ with respect to the aligned case
\cite{na05}. Again, much larger enhancements are possible if the
condition $K_i=\bar K_i$ is relaxed.

%%%%%%%%%%%%%%%%%%%%
\section{Discussion} 
\label{sec:discussion}
In this section we explain some generic features of our results. We
refer here to results that, while depending on the specific flavor
structures, are qualitatively similar in all the temperature regimes
in which flavor effects are important.  For the purposes of this
discussion, it is important the fact that in all the regimes
considered the final $Y_{\Delta_i}$ asymmetries are inversely
proportional to the rates of the corresponding washout processes. This
can be demonstrated along lines similar to those given in Appendix~2
of ref.~\cite{ba00} for $Y_{B-L}$. Note that we have already used this
result extensively for the qualitative discussion in
section~\ref{subs:flavorAsy}. In all the temperature regimes where
flavor effects are important ($M_1\ll 10^{13}~{\rm GeV}$), the washout
rate having the strongest impact on the final value of the asymmetries
is $\gamma_{N_s}$, that is the on-shell piece of the $\Delta L=2$
scattering (the term proportional to $K^0_i\gamma_D$ in the first line
of eq.~(\ref{eqyli})) that has a Boltzmann suppression factor
exp($-z$), similar to the $\Delta L=1$ rates. Hence, the proof given in
ref.~\cite{ba00} for the case of $\Delta L=1$ washout dominance and
small departure from equilibrium, holds also for the cases considered
here, and applies to the single flavor density asymmetries as well.

As is written explicitly in eq.~(\ref{eq:4flavor}), the final values
of the $Y_{\Delta_i}$ are determined by the asymmetry parameters
$\epsilon_1^i$ and by the washout factors $\eta_i$. The washout
factors are related to the various lepton number violating processes
of eq. (\ref{eqyli2}) and hence depend on the factors $K^0_i$ that
control the overall strength of the washouts, as well as on the matrix
coefficients $C^\ell$ and $C^H$ defined in (\ref{AH}). However, the
matrices $C^\ell$ in eqs. (\ref{case3}), (\ref{case4}), (\ref{case5})
and (\ref{case6}) are approximately diagonal and the diagonal terms in
each matrix are not very different from each other (actually, for each
temperature regime, $C^\ell_{ii}$ equals the value of $c_{\ell}$ in
the case that $K^0_i=1$, see \cite{na05}). If we make the crude
approximations of (i) neglecting the off-diagonal elements, and (ii)
taking the diagonal elements equal (that is $C^\ell\propto 1$), and we
also note that the same Higgs asymmetry $y_H$ enters the equation for
the different flavors, we are led to conclude that the relative values
of the $\eta_i$ are determined mainly by the values of the $K^0_i$.
More precisely, given that the amount of flavor-asymmetries surviving
the washout are, to a good approximation, inversely proportional to
the washout rates, the washout factors obey the approximate
proportionality $\eta_i\propto 1/K_i$. This results constitutes the
basis of the approximate expressions for $n_B/s$ in
eq.~(\ref{eq:4eta2}) that allowed us to estimate qualitatively all the
most important effects of flavor.  We rely again on this approximation
in the following discussion of the effects of different flavor
structures and, in particular, in our attempt to understand the
different behaviors that are apparent from the points labeled $(a)$,
$(b)$, $(c)$ and $(d)$ in Fig.~\ref{figure1}.

%%%%%%%%%%%%%%%%%%%%%%%%%%%%%%%%%%%%%%%%%%%%%%%%%%%%%%%%%%%%%%%%%%%%%%%
% PANEL
\begin{figure}[t!]
\vskip-2mm
\hskip-4mm
 \includegraphics[width=6cm,height=8cm,angle=270]{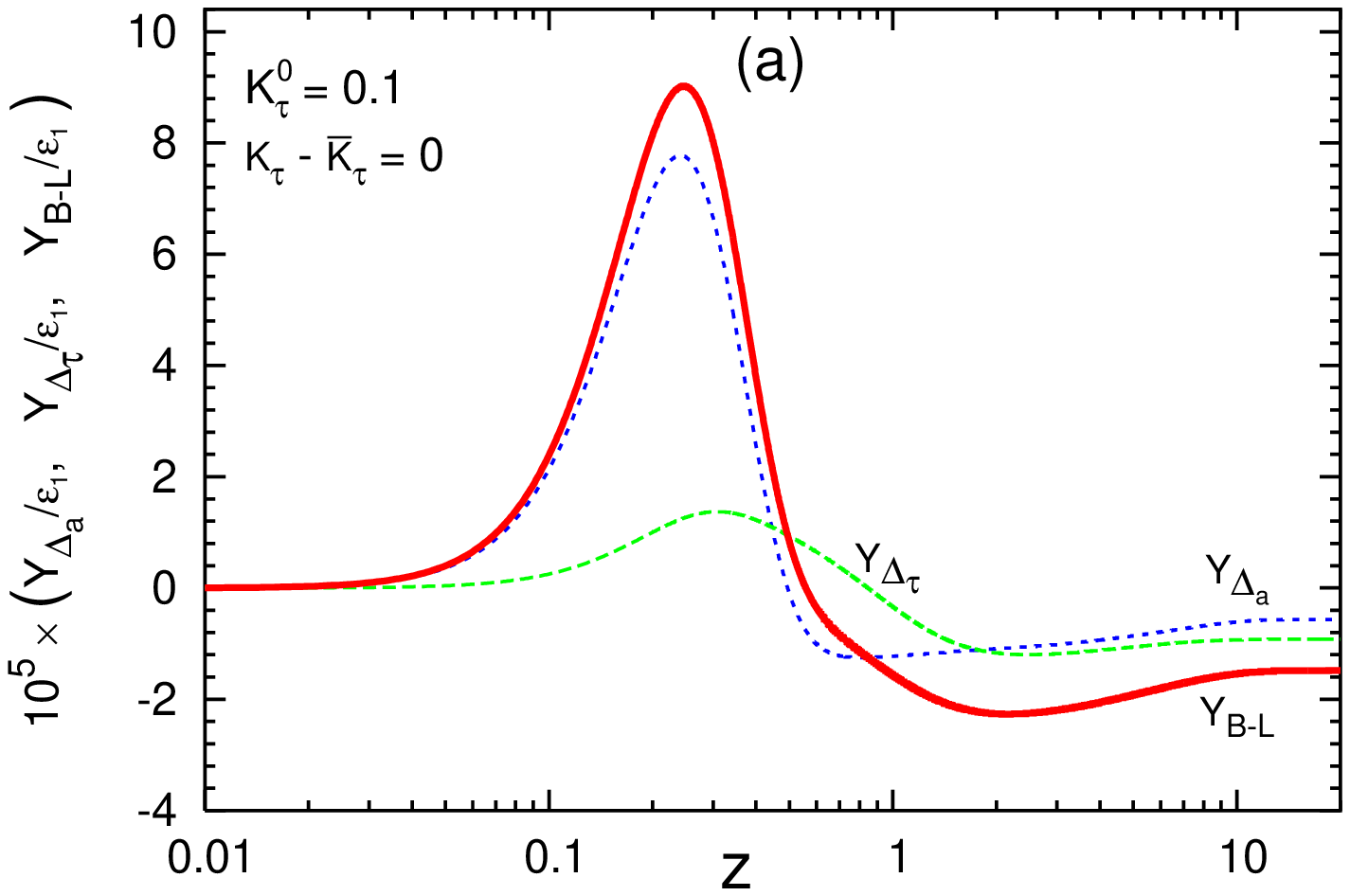}
 \includegraphics[width=6cm,height=8cm,angle=270]{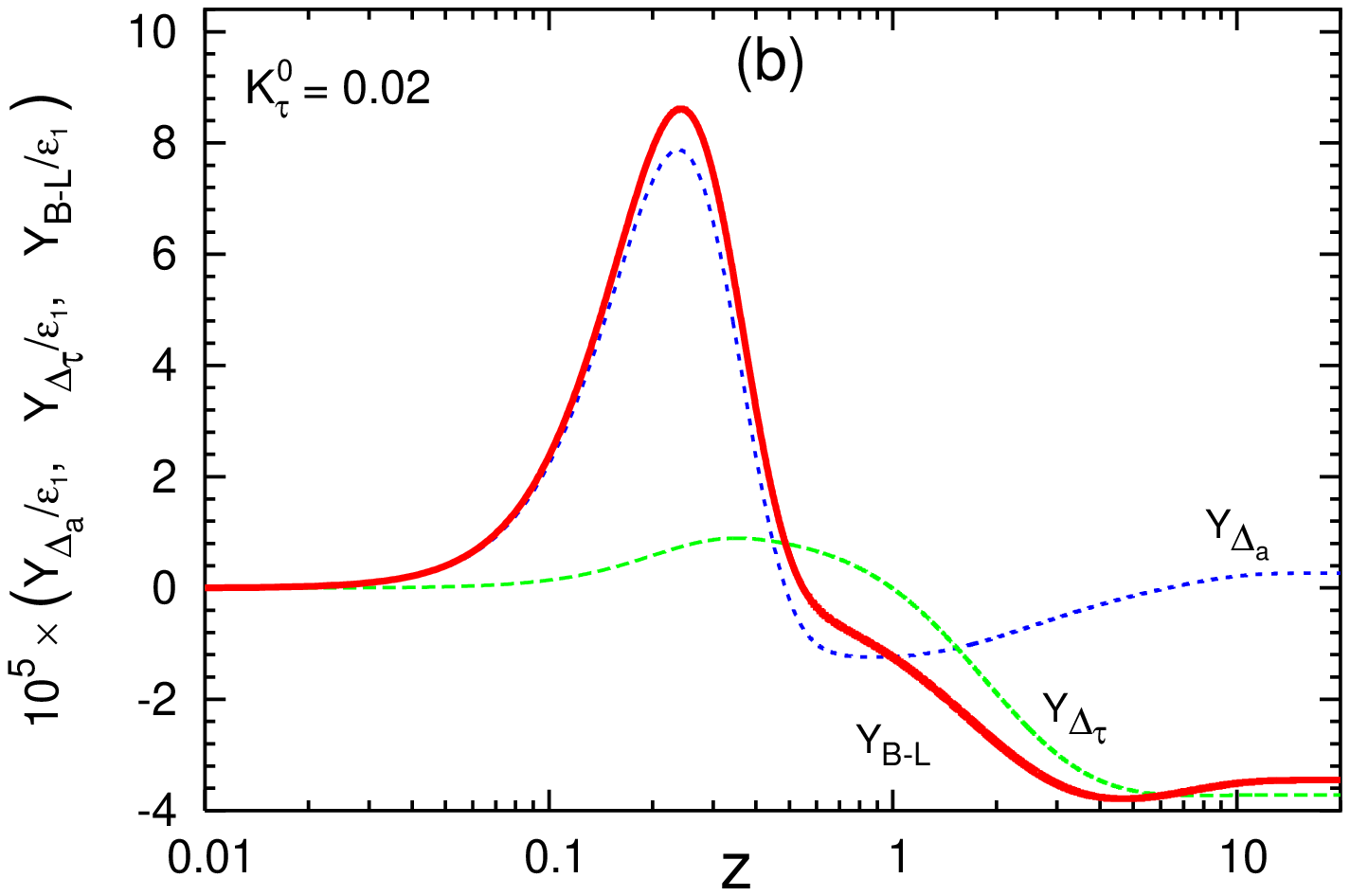}
\vskip-2mm
\hskip-4mm
 \includegraphics[width=6cm,height=8cm,angle=270]{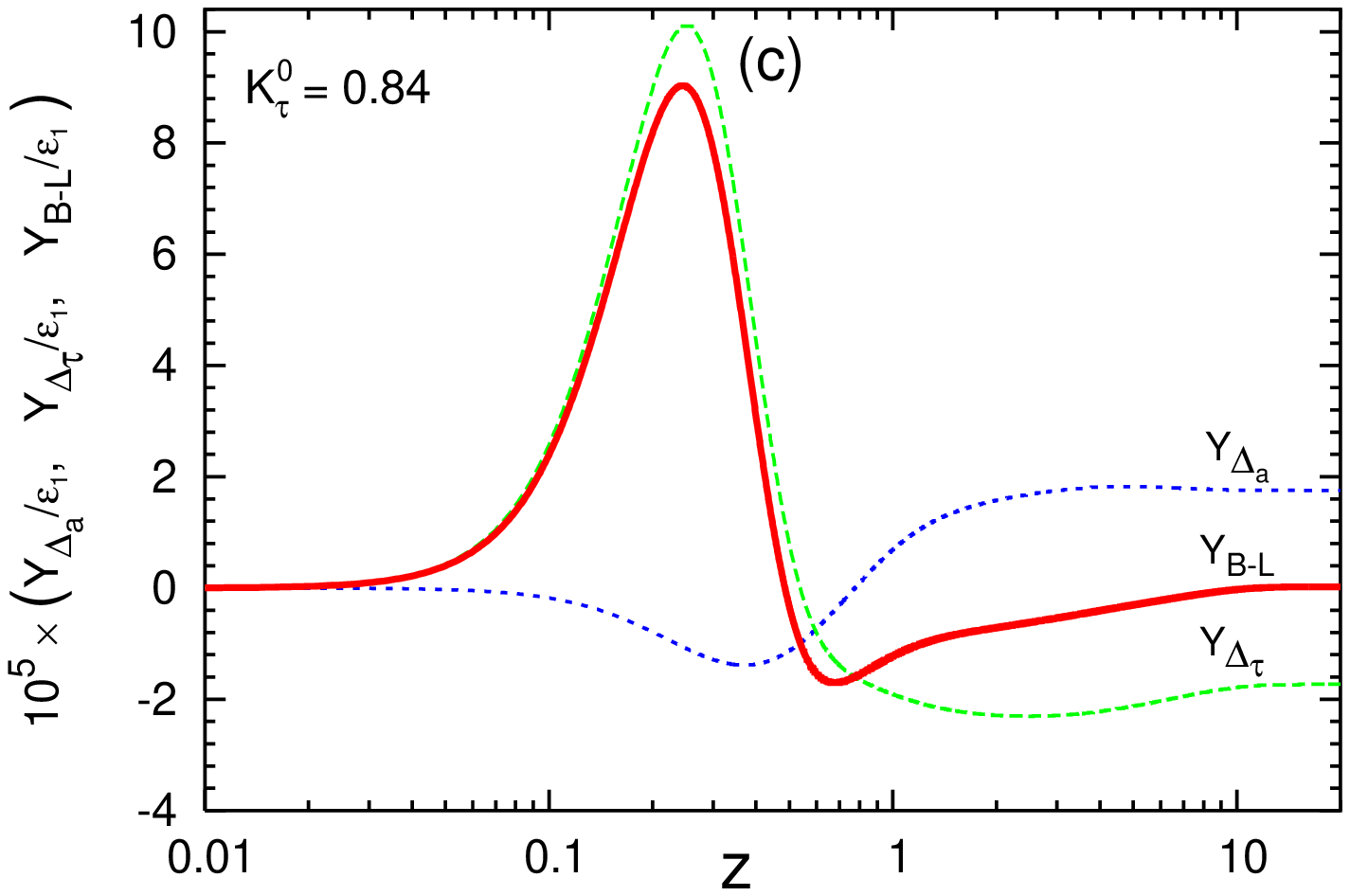}
 \includegraphics[width=6cm,height=8cm,angle=270]{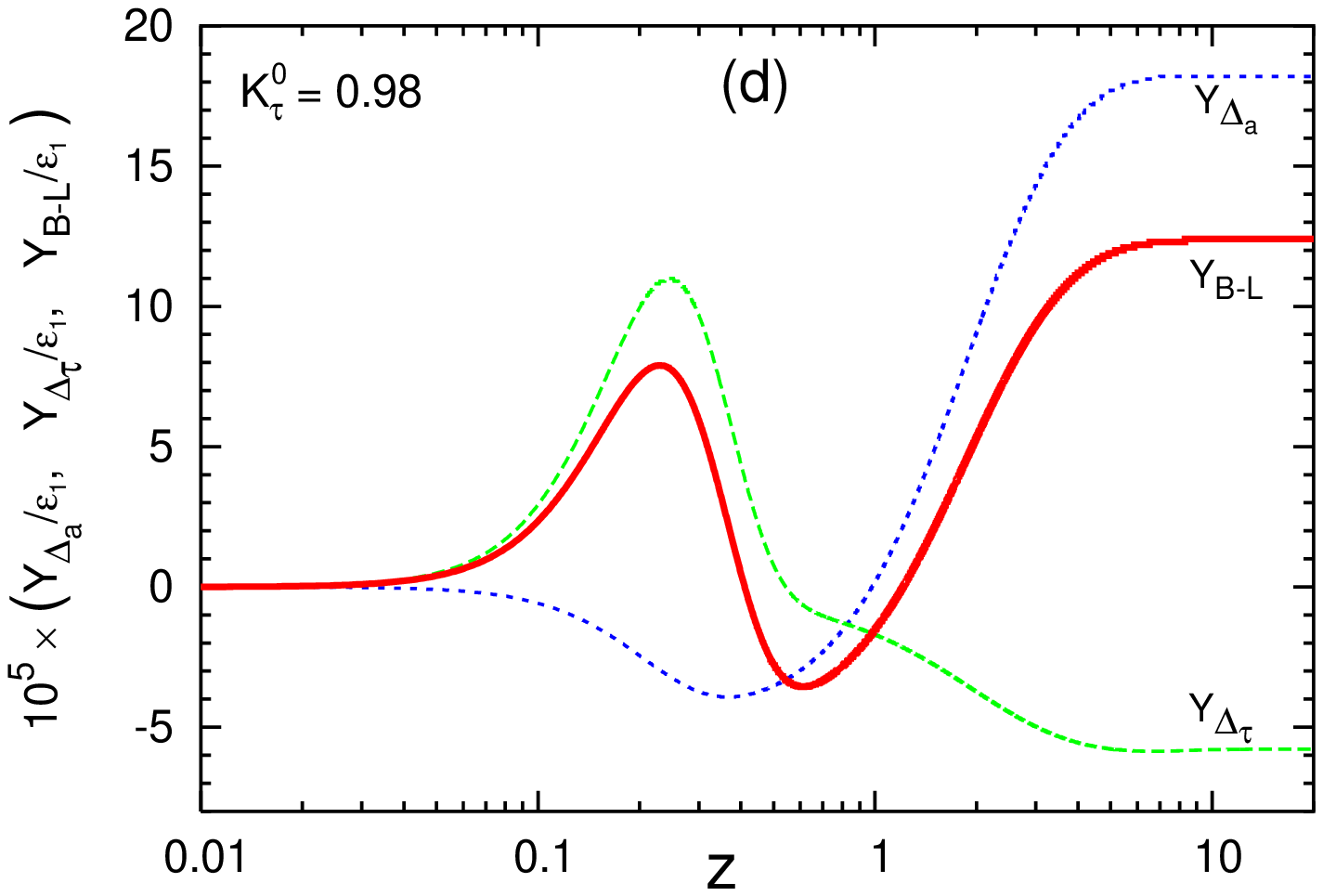}
%
% \caption[]{\baselineskip 12pt
%
\caption[]{ $Y_{\Delta_\tau}/\epsilon_1$ (short-dashed line),
  $Y_{\Delta_a}/\epsilon_1$ (dotted line) and $Y_{B-L}/\epsilon_1$
  (solid line) as a function of $z=M_1/T$ computed with
  $M_1=10^{11}$~GeV and $\tilde m_1=0.06$~eV, in the  temperature
  regime 2.  The panels refer to the labels in fig.~1. The examples
  include both a constrained case, $\Delta K_\tau=0$ ($a$), and
  general cases with $\Delta K_\tau/2\epsilon = \sqrt{K^0_\tau}/4$
  ($b,c,d$). The $K^0_\tau$ value is $(a)$ $0.1$, $(b)$ $0.02$,  $(c)$
  $0.84$ and $(d)$ $0.98$. Note that the vertical scale in $(d)$ is
  doubled with respect to the other three cases.}
\label{figure2}
\end{figure} 
%
%%%%%%%%%%%%%%%%%%%%%%%%%%%%%%%%%%%%%%%%%%%%%%%%%%%%%%%%%%%%%%%%%%%%%%%

$(a)$ This case is defined by the condition $K_i=\bar K_i$ or,
more generally, by the condition $\Delta K_i\ll\epsilon_1\,K^0_i$ for
all relevant $i$.  As we have seen, in this case the final $Y_{B-L}$
shows a sizable enhancement with respect to the aligned cases in the
same temperature regime, and its value is approximately independent of
$K^0_i$. This behavior corresponds to the $\sim n_f$ enhancement
discussed in section~\ref{subs:flavorAsy}: since $K_i=\bar K_i$
implies $\epsilon_1^i=K^0_i \, \epsilon $, it follows that, to first 
approximation, each $Y_{\Delta_i}\propto \epsilon_1^i\,\eta_i$ 
($i=a,\tau$ in regimes (1) and (2), $i=e,\mu,\tau$ in regimes (3) and  
(4)) is independent of $K^0_i$ and, moreover, that the final values of
the $Y_{\Delta_i}$-asymmetries are in general quite similar. The
enhancement of $Y_{B-L}$ by a factor $\sim n_f$ with respect to
the aligned cases \cite{na05} is indeed confirmed by the numerical
results given in Table~\ref{tab:results} ($n_f=2$ in regimes (1) and
(2), $n_f=3$ in regimes (3) and (4)). This case is illustrated in
more detail in Fig.\ref{figure2}$\,a$.  The figure shows the evolution
of $Y_{\Delta_a}$ and $Y_{\Delta_\tau}$ with $z$ in the regime 2, for
$K^0_\tau=0.1$. It is apparent that in spite of the large difference
in the values of $K^0_a$ and $K^0_\tau$, that result in quite
different evolutions for the two asymmetries, their final values are
very similar and approximately equal to $Y_{B-L}/2$.

Note that this result relies in a crucial way on the assumption that
the source terms $\epsilon_1^i$ and the washout rates are both
proportional to $K^0_i$. If this condition is not realized, as in the
general cases when $K\neq \bar K$, more subtle effects become
important in determining the final values of the asymmetries
$Y_{\Delta_i}$ and, as is apparent from Fig.\ref{figure1}, the final
value of $Y_{B-L}$ acquires a rather strong dependence on the values
of $K^0_i$.  Let us now discuss a few examples of this more general
case. 

$(b)$ In this case, $\Delta K_\tau/2$ and $\epsilon_1\,K^0_\tau$ are
of the same order and of the same sign. The details of this case are
illustrated in Fig.\ref{figure2}$\,b$ where we present the behavior of
$Y_{\Delta_\tau}$ and $Y_{\Delta_a}$ for $K^0_\tau=0.02$,
that corresponds to $\Delta K_\tau/2\epsilon = \sqrt{K^0_\tau}/4 \simeq
0.035$. The picture explains rather well the origin of the first peak,
labeled with $(b)$ in Fig.\ref{figure1}. We see that a deviation at
the level of just a few percent from exact alignment ($K^0_\tau=0$) is
sufficient to start populating the asymmetry in the $\tau$ flavor even
if the rate, which is suppressed by $K^0_\tau+ \sqrt{K^0_\tau}/4\sim
1/20$, is rather small. For the first part of the leptogenesis era,
$Y_{\Delta_\tau}$ gives only a minor contribution to the total $B-L$
asymmetry. However, while the washout in the direction
$\ell_a(\perp\ell_\tau)$, that is controlled by $K^0_a\approx 1$,
proceeds with full strength, there is practically no washout for the
$\tau$ component and, eventually, while $Y_{\Delta_a}$ ends up being
almost completely washed out, it is $Y_{\Delta_\tau}$ that determines
the final value of $B-L$ at the end of the leptogenesis era.

$(c)$ This intriguing case corresponds to $\epsilon_1^\tau \eta_\tau
\approx -\epsilon_1^a\eta_a$: The final values of the two leptonic
asymmetries are approximately equal in magnitude but of opposite
signs. This implies that $Y_{B-L}$ can vanish even when lepton flavor
asymmetries are sizable. Eq.~(\ref{eq:4flavor}) helps us to
understand this case in a simple way. Adopting the approximation
$\eta_i\sim \eta/K^0_i$ discussed above we have:
\begin{equation}
\epsilon_1^a\eta_a + \epsilon_1^\tau\eta_\tau\ \approx \ \eta\left[  
2\,\epsilon_1 - \left(\frac{1}{1-K^0_\tau}-\frac{1}{K^0_\tau}\right)
\frac{\Delta K_\tau}{2}\right].
\label{vanish}
 \end{equation}
With our particular ansatz, $\Delta K_\tau/{2\epsilon_1}=
\sqrt{K^0_\tau}/4 $, the r.h.s. of eq.~(\ref{vanish}) vanishes for
$K^0_\tau\simeq 0.9$, in qualitative agreement with the point labeled
with $(c)$ in Fig.~\ref{figure1}. The evolution with $z$ of the
flavor asymmetries is illustrated in Fig.\ref{figure2}$\,c$. During
most of the leptogenesis era, $Y_{B-L}$ is rather large. However, in
the end its value is driven to zero, even if the final values of the
two lepton asymmetries remain quite sizable (about twice larger
than case~$(a)$).  For larger values of $K^0_\tau$ the total
asymmetry $Y_{B-L}$ crosses zero and changes sign, and we enter a
different regime to be discussed next. 

$(d)$ In this example we have $\Delta K_a/2\epsilon_1\simeq -0.25$ and
$K_a^0=0.02$, so that $\Delta K_a/2\epsilon_1$ is much larger in
absolute value than $K^0_a$, and is negative. This situation means
that $\epsilon_1^a$ has a sign opposite to both $\epsilon_1^\tau$ and
$\epsilon_1$. This case is illustrated in Fig.~\ref{figure2}$\,d$, in
which the vertical scale is doubled with respect to the previous three
cases. Initially, a large density-asymmetry starts being built in the
$\tau$ flavor, due to the large value $\epsilon_1^\tau/\epsilon_1 \sim
1.23$. However, its growth is kept under control by the washout
effects that, with $K^0_\tau  \approx 1$, are unsuppressed. In
contrast, for $Y_{\Delta_a}$ the washout effects are strongly
suppressed by $K^0_a\sim 0.02$ and eventually this asymmetry largely
prevails, driving $Y_{B-L}$ to  values several times larger than in
all the previous cases, and having opposite sign. Notice that in this
case the difference in sign between $Y_{\Delta_a}$ and
$Y_{\Delta_\tau}$ gives large cancellations between the respective
contributions to the asymmetry  $y_H$, and this implies that the Higgs
washout effects are accordingly suppressed.

%%%%%%%%%%%%%%%%%%%%%%%%%%%
\section{Conclusions}
\label{sec:conclusions}
We have shown that the effects of flavor can have dramatic
consequences in leptogenesis scenarios. This occurs due to the way in
which charged lepton Yukawa interactions in thermal equilibrium affect
the flavor composition of the leptonic density-asymmetries, that
determine the washout processes. For these effects to be significant,
at least one leptonic Yukawa interaction ($h_\tau$) must be in
equilibrium. This happens if $M_1$, the lightest heavy neutrino mass
that determines the temperature at which leptogenesis takes place, is
light enough: $M_1<10^{13}$~GeV. Moreover, for the flavor effects to
have an impact, one needs to be in the strong washout regime:
$\tilde m_1>2\times 10^{-3}$~eV.

The main consequence of the flavor effects is that, in generic flavor
non-aligned cases with strong washouts, the final asymmetry is
typically enhanced by a factor $n_f$. We have $n_f=2$ when just
$h_\tau$ is in equilibrium, i.e. for $10^{9}\ {\rm GeV}<M_1<10^{13}\
{\rm GeV}$, and $n_f=3$ when also $h_\mu$ is in equilibrium, i.e. for
$M_1<10^{9}\ {\rm GeV}$.

In addition to the total asymmetry $\epsilon_1$ associated to $N_1$
decays, the individual asymmetries $\epsilon_1^j$ can play a crucial
role. These flavor asymmetries introduce a qualitatively new
ingredient, $\Delta K_j$, which is the contribution associated to the
fact that the leptons and antileptons produced in $N_1$ decays are, in
general, not $CP$ conjugates of each other. When these new
contributions are non-negligible, much larger enhancements become
possible, especially when at least one lepton flavor $\ell_j$ is
weakly coupled to the decaying $N_1$ (typically this means
$K^0_j\sim \eta\ll 1$). Since the signs of $\Delta K_j$ and of
$\epsilon_1$ need not be the same, the final asymmetry can have a sign
unrelated to that of $\epsilon_1$. Actually, successful leptogenesis
is possible even with $\epsilon_1=0$.  Scenarios in which
$\epsilon_1=0$ while $\epsilon^j_1\neq 0$ entail the possibility
that the phases in the light neutrino mixing matrix $U$ are 
the only source of $CP$ violation. 

%%%%%%%%%%%%%%%%%%%%%%%%%%
\section*{Acknowledgments}
We thank the authors of ref. \cite{ab06} (and, in particular, Sacha
Davidson) for making their related work available to us prior to
publication, and for suggesting a simultaneous submission to the
arXiv. We are grateful to Sacha Davidson and Marta Losada for useful
comments. E.N. acknowledges D. Aristizabal for his help in
cross-checking some of the computations.  Work supported in part by
ANPCyT and Fundaci\'on Antorchas, Argentina, by the Istituto Nazionale
di Fisica Nucleare (INFN), Italy, and by Colciencias in Colombia under
contract 1115-05-13809.  The work of Y.N. is supported by the Israel
Science Foundation founded by the Israel Academy of Sciences and
Humanities, by EEC RTN contract HPRN-CT-00292-2002, and by the United
States-Israel Binational Science Foundation (BSF), Jerusalem, Israel.

%%%% ======================================================== 


\vspace{-.3cm}

\begin{thebibliography}{99}

\vspace{-.2cm}


\bibitem{fu86} M.~Fukugita and T.~Yanagida,
  %``Baryogenesis Without Grand Unification,''
  Phys.\ Lett.\ B {\bf 174}, 45 (1986).
  %%CITATION = PHLTA,B174,45;%%

\bibitem{lu92} M.~A.~Luty,
  %``Baryogenesis via leptogenesis,''
  Phys.\ Rev.\ D {\bf 45}, 455 (1992).
  %%CITATION = PHRVA,D45,455;%%

\bibitem{seesaw} P. Minkowski, {\it Phys. Lett.} B {\bf 67} 421 (1977); 
T. Yanagida, in {\it Proc. of Workshop on Unified Theory and Baryon
number in the Universe}, eds. O. Sawada and A. Sugamoto, KEK, Tsukuba, (1979) p.95;
M. Gell-Mann, P. Ramond and R. Slansky,  in {\it Supergravity}, eds P. 
van Niewenhuizen and D. Z. Freedman (North Holland, Amsterdam 1980) p.315;
P. Ramond, {\it  Sanibel talk}, retroprinted as hep-ph/9809459;
S. L. Glashow, in {\it Quarks and Leptons}, Carg\`ese lectures, eds M. L\'evy,
(Plenum, 1980, New York) p. 707;
R. N. Mohapatra and G. Senjanovi\'c, {\it Phys. Rev. Lett.} {\bf 44}, 912 (1980).

%\cite{Sakharov:1967dj}
%\bibitem{Sakharov:1967dj}
\bibitem{sa67}
  A.~D.~Sakharov,
  %``Violation Of CP Invariance, C Asymmetry, And Baryon Asymmetry Of The
  %Universe,''
  Pisma Zh.\ Eksp.\ Teor.\ Fiz.\  {\bf 5} (1967) 32
  [JETP Lett.\  {\bf 5} (1967\ SOPUA,34,392-393.1991\ UFNAA,161,61-64.1991) 24].
  %%CITATION = ZFPRA,5,32;%%

%\cite{Kuzmin:1985mm}
%\bibitem{Kuzmin:1985mm}
\bibitem{kuz85}
  V.~A.~Kuzmin, V.~A.~Rubakov and M.~E.~Shaposhnikov,
  %``On The Anomalous Electroweak Baryon Number Nonconservation In The Early
  %Universe,''
  Phys.\ Lett.\ B {\bf 155}, 36 (1985).
  %%CITATION = PHLTA,B155,36;%%


\bibitem{BBP0204}
%\cite{Buchmuller:2004tu}
% \bibitem{Buchmuller:2004tu}
  W.~Buchmuller, P.~Di Bari and M.~Plumacher,
  %``Some aspects of thermal leptogenesis,''
  New J.\ Phys.\  {\bf 6}, 105 (2004) 
  [arXiv:hep-ph/0406014]; 
  %%CITATION = HEP-PH 0406014;%%
%
%\cite{Buchmuller:2004nz}
% \bibitem{Buchmuller:2004nz}
%  W.~Buchmuller, P.~Di Bari and M.~Plumacher,
  %``Leptogenesis for pedestrians,''
  Annals Phys.\  {\bf 315}, 305  (2005) 
  [arXiv:hep-ph/0401240];
  %%CITATION = HEP-PH 0401240;%%
%
%\cite{Buchmuller:2002jk}
% \bibitem{Buchmuller:2002jk}
%  W.~Buchmuller, P.~Di Bari and M.~Plumacher,
  %``A bound on neutrino masses from baryogenesis,''
  Phys.\ Lett.\ B {\bf 547},  128 (2002)
  [arXiv:hep-ph/0209301];
  %%CITATION = HEP-PH 0209301;%%
%
%\cite{Buchmuller:2002rq}
% \bibitem{Buchmuller:2002rq}
%  W.~Buchmuller, P.~Di Bari and M.~Plumacher,
  %``Cosmic microwave background, matter-antimatter asymmetry and neutrino
  %masses,''
  Nucl.\ Phys.\ B {\bf 643}, 367  (2002) 
  [arXiv:hep-ph/0205349].
  %%CITATION = HEP-PH 0205349;%%

\bibitem{BP9900}
%\cite{Buchmuller:2000as}
% \bibitem{Buchmuller:2000as}
  W.~Buchmuller and M.~Plumacher,
  %``Neutrino masses and the baryon asymmetry,''
  Int.\ J.\ Mod.\ Phys.\ A {\bf 15}, 5047  (2000) 
  [arXiv:hep-ph/0007176];
  %%CITATION = HEP-PH 0007176;%%
%
%\cite{Buchmuller:1999cu}
% \bibitem{Buchmuller:1999cu}
%  W.~Buchmuller and M.~Plumacher,
  %``Matter antimatter asymmetry and neutrino properties,''
  Phys.\ Rept.\  {\bf 320}, 329 (1999) 
  [arXiv:hep-ph/9904310].
  %%CITATION = HEP-PH 9904310;%%

\bibitem{pi04} A.~Pilaftsis and T.~E.~J.~Underwood,
  %``Resonant leptogenesis,''
  Nucl.\ Phys.\ B {\bf 692}, 303 (2004)
  [arXiv:hep-ph/0309342].
  %%CITATION = HEP-PH 0309342;%%

%\cite{Pilaftsis:2005rv}
% \bibitem{Pilaftsis:2005rv}
 \bibitem{pi05}
  A.~Pilaftsis and T.~E.~J.~Underwood,
  %``Electroweak-scale resonant leptogenesis,''
  Phys.\ Rev.\ D {\bf 72}, 113001 (2005) 
  [arXiv:hep-ph/0506107].
  %%CITATION = HEP-PH 0506107;%%

%\cite{Hambye:2003rt}
% \bibitem{Hambye:2003rt}
\bibitem{ha04}
  T.~Hambye {\it et al.},
  %Y.~Lin, A.~Notari, M.~Papucci and A.~Strumia,
  %``Constraints on neutrino masses from leptogenesis models,''
  Nucl.\ Phys.\ B {\bf 695}, 169 (2004)
  [arXiv:hep-ph/0312203].
  %%CITATION = HEP-PH 0312203;%%


\bibitem{gi04} G.~F.~Giudice {\it et al.},
  %A.~Notari, M.~Raidal, A.~Riotto and A.~Strumia,
  %``Towards a complete theory of thermal leptogenesis in the SM and MSSM,''
  Nucl.\ Phys.\ B {\bf 685}, 89 (2004)
  [arXiv:hep-ph/0310123].
  %%CITATION = HEP-PH 0310123;%%


\bibitem{bu01} W.~Buchmuller and M.~Plumacher,
  %``Spectator processes and baryogenesis,''
  Phys.\ Lett.\ B {\bf 511}, 74 (2001)
  [arXiv:hep-ph/0104189].
  %%CITATION = HEP-PH 0104189;%%

\bibitem{na05}
%\cite{Nardi:2005hs}
%\bibitem{Nardi:2005hs}
  E.~Nardi, Y.~Nir, J.~Racker and E.~Roulet,
  %``On Higgs and sphaleron effects during the leptogenesis era,''
  JHEP {\bf 0601}, 068 (2006) [arXiv:hep-ph/0512052].
  %%CITATION = HEP-PH 0512052;%%

\bibitem{ba00}
  R.~Barbieri, P.~Creminelli, A.~Strumia and N.~Tetradis,
  %``Baryogenesis through leptogenesis,''
  Nucl.\ Phys.\ B {\bf 575}, 61 (2000)
  (for the updated version of this paper see [arXiv:hep-ph/9911315]).
  %%CITATION = HEP-PH 9911315;%%

%\cite{Endoh:2003mz}
%\bibitem{Endoh:2003mz}
\bibitem{en03}
  T.~Endoh, T.~Morozumi and Z.~h.~Xiong,
  %``Primordial lepton family asymmetries in seesaw model,''
  Prog.\ Theor.\ Phys.\  {\bf 111}, 123 (2004)
  [arXiv:hep-ph/0308276].
  %%CITATION = HEP-PH 0308276;%%

%\cite{Fujihara:2005pv}
% \bibitem{Fujihara:2005pv}
\bibitem{fu05}
  T.~Fujihara, S.~Kaneko, S.~Kang, D.~Kimura, T.~Morozumi and M.~Tanimoto,
  %``Cosmological family asymmetry and CP violation,''
  Phys.\ Rev.\ D {\bf 72}, 016006 (2005)
  [arXiv:hep-ph/0505076].
  %%CITATION = HEP-PH 0505076;%%


%\cite{DiBari:2005st}
% \bibitem{DiBari:2005st}
\bibitem{diba05}
  P.~Di Bari,
  %``Seesaw geometry and leptogenesis,''
  Nucl.\ Phys.\ B {\bf 727}, 318 (2005)
  [arXiv:hep-ph/0502082].
  %%CITATION = HEP-PH 0502082;%%


%\cite{Vives:2005ra}
%\bibitem{Vives:2005ra}
\bibitem{vi05}
  O.~Vives,
  %``Flavoured leptogenesis: A successful thermal leptogenesis with N(1) mass
  %below 10**8-GeV,''
  arXiv:hep-ph/0512160.
  %%CITATION = HEP-PH 0512160;%%


  \bibitem{co96} L.~Covi, E.~Roulet and F.~Vissani,
  %``CP violating decays in leptogenesis scenarios,''
  Phys.\ Lett.\ B {\bf 384}, 169 (1996)
  [arXiv:hep-ph/9605319].
  %%CITATION = HEP-PH 9605319;%%

%\cite{Casas:2001sr}
% \bibitem{Casas:2001sr}
\bibitem{ca01}
  J.~A.~Casas and A.~Ibarra,
  %``Oscillating neutrinos and mu $\to$ e, gamma,''
  Nucl.\ Phys.\ B {\bf 618} (2001) 171
  [arXiv:hep-ph/0103065].
  %%CITATION = HEP-PH 0103065;%%

\bibitem{ab06}  
 A. Abada, S. Davidson, F-X. Josse-Michaux, M. Losada and A. Riotto, 
 arXiv:hep-ph/0601083.

\bibitem{ha90} J.~A.~Harvey and M.~S.~Turner,
  %``Cosmological Baryon And Lepton Number In The Presence Of Electroweak
  %Fermion Number Violation,''
  Phys.\ Rev.\ D {\bf 42}, 3344 (1990).
  %%CITATION = PHRVA,D42,3344;%%

\bibitem{la00} M.~Laine and M.~E.~Shaposhnikov,
  %``A remark on sphaleron erasure of baryon asymmetry,''
  Phys.\ Rev.\ D {\bf 61}, 117302 (2000)
  [arXiv:hep-ph/9911473].
  %%CITATION = HEP-PH 9911473;%%

\bibitem{be03} L.~Bento,
  %``Sphaleron relaxation temperatures,''
  JCAP {\bf 0311}, 002 (2003)
  [arXiv:hep-ph/0304263].
  %%CITATION = HEP-PH 0304263;%%

\bibitem{mo97} G.~D.~Moore,
  %``Computing the strong sphaleron rate,''
  Phys.\ Lett.\ B {\bf 412}, 359 (1997)
  [arXiv:hep-ph/9705248].
  %%CITATION = HEP-PH 9705248;%%

\bibitem{mo92} R.~N.~Mohapatra and X.~m.~Zhang,
  %``QCD sphalerons at high temperature and baryogenesis at electroweak scale,''
  Phys.\ Rev.\ D {\bf 45}, 2699 (1992).
  %%CITATION = PHRVA,D45,2699;%%

\bibitem{ar98} P.~Arnold, D.~T.~Son and L.~G.~Yaffe,
  %``Hot B violation, color conductivity, and log(1/alpha) effects,''
  Phys.\ Rev.\ D {\bf 59}, 105020 (1999)
  [arXiv:hep-ph/9810216].
  %%CITATION = HEP-PH 9810216;%%

\bibitem{bodeker98} D. B\"odeker,
  %``On the effective dynamics of soft non-abelian gauge fields at finite
  %temperature,''
  Phys.\ Lett.\ B {\bf 426}, 351 (1998)
  [arXiv:hep-ph/9801430].
  %%CITATION = HEP-PH 9801430;%%

\bibitem{ar97} P.~Arnold, D.~Son and L.~G.~Yaffe,
  %``The hot baryon violation rate is O(alpha(w)**5 T**4),''
  Phys.\ Rev.\ D {\bf 55}, 6264 (1997)
  [arXiv:hep-ph/9609481].
  %%CITATION = HEP-PH 9609481;%%


\end{thebibliography}
\end{document}